\documentclass[fleqn,usenatbib,useAMS]{mnras}

\usepackage{graphicx}	% Including figure files
\usepackage{amsmath}	% Advanced maths commands
\usepackage{amssymb}	% Extra maths symbols
\usepackage{multicol}        % Multi-column entries in tables
\usepackage{bm}		% Bold maths symbols, including upright Greek
\usepackage{pdflscape}	% Landscape pages

\usepackage[T1]{fontenc}
\usepackage{ae,aecompl}

\usepackage{newtxtext,newtxmath}
\title[Galaxy Mergers in UNIONS]{Galaxy Mergers in UNIONS -- I: A Simulation-driven Hybrid Deep Learning Ensemble for Pure Galaxy Merger Classification}
\author[L. Ferreira et al.]{
Leonardo Ferreira$^{1}$\thanks{Contact e-mail: \href{lferreira@uvic.ca}{lferreira@uvic.ca}},
    Robert W. Bickley$^{1}$,
    Sara L. Ellison$^{1}$,
    David R. Patton$^{2}$,
    Shoshannah Byrne-Mamahit$^{1}$,
    \newauthor
    \ Scott Wilkinson$^{1}$,
    Connor Bottrell$^{3}$,  
    Sébastien Fabbro$^{4}$.
    Stephen D. J. Gwyn$^{4}$,
    Alan McConnachie$^{4}$
   \\
$^{1}$School of Physics and Astronomy, University of Victoria, Victoria, BC, Canada\\
$^{2}$Department of Physics and Astronomy, Trent University, 1600 West Bank Drive, Peterborough, ON K9L 0G2, Canada \\
$^{3}$International Centre for Radio Astronomy Research, University of Western Australia, 35 Stirling Hwy, Crawley, WA 6009, Australia \\
$^{4}$NRC Herzberg Astronomy and Astrophysics, 5071 West Saanich Road, Victoria, BC, V9E2E7, Canada}

\date{Accepted 2024 July 22. Received 2024 July 22; in original form 2024 April 03}

\pubyear{2024}

\begin{document}
\label{firstpage}
\pagerange{\pageref{firstpage}--\pageref{lastpage}}
\maketitle

\begin{abstract}
Merging and interactions can radically transform galaxies. However, identifying these events based solely on structure is challenging as the status of observed mergers is not easily accessible. Fortunately, cosmological simulations are now able to produce more realistic galaxy morphologies, allowing us to directly trace galaxy transformation throughout the merger sequence. To advance the potential of observational analysis closer to what is possible in simulations, we introduce a supervised deep learning Convolutional Neural Network (CNN) and Vision Transformer (ViT) hybrid framework, \textsc{Mummi} (\textbf{MU}lti \textbf{M}odel \textbf{M}erger \textbf{I}dentifier). \textsc{Mummi} is trained on realism-added synthetic data from IllustrisTNG100-1, and is comprised of a multi-step ensemble of models to identify mergers and non-mergers, and to subsequently classify the mergers as interacting pairs or post-mergers. To train this ensemble of models, we generate a large imaging dataset of 6.4 million images targeting UNIONS with \texttt{RealSimCFIS}. We show that \textsc{Mummi} offers a significant improvement over many previous machine learning classifiers, achieving $95\%$ pure classifications even at Gyr long timescales when using a jury-based decision making process, mitigating class imbalance issues that arise when identifying real galaxy mergers from $z=0$ to $0.3$. Additionally, we can divide the identified mergers into pairs and post-mergers at $96\%$ success rate. We drastically decrease the false positive rate in galaxy merger samples by $75\%$. By applying \textsc{Mummi} to the UNIONS DR5-SDSS DR7 overlap, we report a catalog of $13,448$ high confidence galaxy merger candidates. Finally, we demonstrate that \textsc{Mummi} produces powerful representations solely using supervised learning, which can be used to bridge galaxy morphologies in simulations and observations. 
\end{abstract}


%%%%%%%%%%%%%%%%%%%%%%%%%%%%%%%%%%%%%%%%%%%%%%%%%%

%%%%%%%%%%%%%%%%% BODY OF PAPER %%%%%%%%%%%%%%%%%%

% The MNRAS class isn't designed to include a table of contents, but for this document one is useful.
% I therefore have to do some kludging to make it work without masses of blank space.
\begingroup
\let\clearpage\relax
\endgroup
\newpage

\section{Introduction}

When the distant nature of extra-galactic nebulae was first revealed,  systematic cataloguing and classification efforts put little focus on highly irregular and disturbed systems. This was mostly due to their rare occurrence in the local universe \citep{Sandage2005}. It was not until \cite{TandT1972} conducted the first numerical simulations of merging disc galaxies that the significant role of merging events in galaxy transformation was demonstrated. This insight put galaxy mergers as a central piece in the puzzle of galaxy formation and evolution. Moreover, these early studies unequivocally showed that merging was the primary mechanism producing highly asymmetric, tidal, and disruptive features \citep{Toomre1977, conselice2003, Patton2005, Casteels2014}.  %OK quality

The current widely adopted cosmological model  -- the  Dark Energy and Cold Dark Matter ($\rm \Lambda CDM$) model -- also emphasizes the important role of galaxy mergers   \citep{Planck2018}. Small-scale perturbations in an expanding Big Bang universe seed the formation of the first dark matter haloes, which then foster the gravitational collapse of baryonic matter into their gravitational wells. The presence of these fluctuations enables the clustering of matter and thus the formation of the first stars and galaxies. However, to produce the cosmic landscape we observe in the local universe, these distant infant galaxies have to coalesce together to form larger, more massive systems \citep{Bluementhal1984}. Undeniably, the occurrence of merging galaxies is a by-product of the Universe's hierarchical assembly  \citep{Duncan2019}. Galaxies grow not only by forming stars from their available gas supply  \citep{Madau2014} but also by coalescing together to form larger systems \citep{Patton2002, Duncan2019}. Tracing this assembly through merging galaxies then uncovers how galaxies transform from their seedlings in the early universe to the systems we see today. 

%This also offers a perspective on the evolution of the Universe itself \citep{Hopkins2008, Conselice2014etal}. %OK might be too repetitive

%modern view on galaxy impacts on galaxy evolution
In addition to their importance in the build-up of stellar mass, galaxy mergers have been both predicted and observed to affect several other internal physical processes: mergers can trigger active galactic nuclei (AGNs) by channeling material into central supermassive black holes \citep{ellison2011, Satyapal2014, Ellison2019, Byrne-Mamahit2023, Li2023, Byrne-MamahitB, Bickley2023, Bickley2024}; mergers can enhance star formation by replenishing and mixing the gas reservoirs of galaxies \citep{scudder2012, patton2013, Violino2018, Garay-Solis2023}; or as a quenching mechanism for star formation \citep{Springel2005, Hopkins2008, ellison2022, Wilkinson2022}, either by driving gas out of the galaxy or by adding energy to the system, which can inhibit gas cooling. On cosmological scales, the merging of low mass end dwarf galaxies can be one of the explanations for the existence of intermediate mass black holes \citep{Conselice2020}. Furthermore, understanding galaxy merger rates across cosmic time is crucial for constraining the modeling of the recently discovered gravitational wave background signal from NANOGrav  \citep{NANOGravSMHB}, as galaxy merger rates are directly related to the frequency of supermassive black hole mergers.  % This might be too much for the paper (the second paragraph)

Unfortunately, galaxy merging has a complex observability, as some signatures are short-lived and highly dependent on the physical configuration of the orbits of the system \citep{Lotz2008, Whitney2021, Wilkinson2024, Patton2024}, hence identifying such events is a challenging task. A plethora of classification systems exist, from visual identification to completely automated deep learning techniques \citep{Conselice2014Review, Huertas-Company2023}. Some methods are specifically designed to identify mergers at certain stages, like the pair statistics approach \citep{Patton2000, Ellison2008, Man2016, Mantha2018, Duncan2019} where galaxies are selected based on their likelihood of merging in the future from a initial stage. Others are optimized for the post-coalescence stage, such as the shape asymmetry index \citep{Pawlik2016, Wilkinson2022} that focuses on disturbances to the outline shape of post-mergers. Conversely, more general methods like the CAS system  \citep[Concentration, Asymmetry, Smoothness,][]{Schade1995, Abraham1996, Bershady2000, Conselice2000, Conselice2003a, Conselice2003b} and the G-$M_{20}$ system \citep[Gini coefficient, Second order moment of light,][]{Lotz2004, Snyder2015, Vicente2019, Rose2023}, are capable of detecting multiple merger stages, but their relationship with the merger timescale is also convoluted \citep{Lotz2008, Snyder2019, Whitney2021, Conselice2022}. Strikingly, \citet{Wilkinson2024} demonstrates that these methods do not recover the majority of mergers in galaxy samples.

A popular contemporary approach to merger identification is the use of supervised deep learning methods, where artificial neural networks (ANNs) are trained to reproduce a set of merger classifications through an iterative optimization procedure called stochastic gradient descent (SGD). These methods have been shown to achieve superhuman performance in classification tasks  \citep{DanNet2012, Bickley2021}, and excel in astronomical applications, including morphological classification and merger identification  \citep[for a review, see][]{Huertas-Company2023}. Among these techniques, to date, Convolutional Neural Networks (CNNs) have proven most successful \citep{Dieleman2015, Huertas-Company2015, Huertas-Company2018, Huertas-Company2019, Walmsley2022}. They learn the best discriminatory visual features from the dataset for the task at hand during training, completely bypassing the requirement for careful feature engineering, previously necessary in other learning algorithms such as Random Forests (RFs) \citep{Snyder2019, Yu-Yen2022} and Linear Discriminant Analysis (LDA) \citep{Ferrari2015, Ferreira2018, Nevin2019, Nevin2023}. However, all these practices require pristine pre-labeled data; biases and spurious labels can significantly impact the final model's performance. Additionally, the size of the training dataset is equally important, with the number of examples during training at least on the same scale as the number of parameters being trained.

For merger identification specifically, a promising approach trains these deep learning models with mock observations from large box cosmological simulations \citep{Pearson2019, Ferreira2020, Bickley2021, Bottrell2022b, Ferreira2022}. This approach circumvents the challenge of obtaining real, unambiguous merger observations for data labeling, as the full evolution of a merger event in the simulation is accessible through merger trees. However, the effectiveness of this procedure lies in the quality of the mock observation -- specifically, their realism and whether the simulation prescriptions have any significant flaws that could result in unrealistic galaxy and interaction features  \citep{Bottrell2019, Alexandra2020, Alexandra2021}.

Recently, the approach of using simulation-driven CNN models has achieved state-of-the-art results in identifying galaxy mergers both in the local and distant universe. For instance, \citet{Bickley2021} achieves $\sim 85\%$ accuracy when using a simulation-driven CNN model in classifying post-mergers in the Ultraviolet Near Infrared Optical Northern Survey (UNIONS) r-band images using a simulation-driven CNN model trained with mock observations from IllustrisTNG \citep{Pillepich2018, Nelson2018}. These classifications demonstrate the impact of merging in star formation rates of post-mergers \citep{Bickley2022} and capture the merger-AGN connection \citep{Bickley2023}. At earlier epochs, \citet{Ferreira2020} employed  a similar approach targeting the Cosmic Assembly Near-infrared Deep Extragalactic Survey (CANDELS) data \citep{Grogin2011, Koekomoer2011}. They measured galaxy merger rates from redshifts $z=0.5$ to 3 that are consistent with previous methods for the first time. Similar performance have been achieve in other works \citep{Pearson2019, Wang2020, Bottrell2022, Ferreira2022, Walmsley2023, Margalef-Bentabol2024}. Despite the achieved performance -- over 80\% in some cases -- the rarity of merger galaxies compared to regular galaxies leads to a high rate of false positives \citep{PearsonEAS}.  This necessitates an additional round of visual classifications to vet the final sample \citep{Bickley2021, Bottrell2022b}. This will likely continue to be the case unless classification models are able to produce a false positive rate that is lower than the intrinsic occurrence of mergers in the universe\footnote{The issue of high contamination is not unique to machine learning methods, and has always been an issue, even with traditional methods.  It is simply a result of Bayes theorem \citep[see Fig.1 in][]{Bottrell2022b}}.

To further advance in the direction of more accurate and pure methods, we introduce \textsc{Mummi}  (\textbf{MU}lti \textbf{M}odel \textbf{M}erger \textbf{I}dentifier), an ensemble of deep learning models that achieves $~95\%$ accuracy when classifying recent mergers, and up to $~80\%$ for mergers on billion year long timescales. \textsc{Mummi} incorporates three novel approaches to galaxy merger classification. First, we explore Vision Transformers as alternatives to CNN-based architectures \citep{SwinT2021} due to its global awareness for feature extraction. Second, we combine CNNs and Vision Transformers together to form a large ensemble of diverse models. Finally, we use a jury voting system that drives our training approach and improves the decision-make process by forming consensus within this ensemble of models. Based on a merger sample drawn from the TNG100-1 simulation, with observational realism that replicates the Ultraviolet Near Infrared Optical Northern Survey (UNIONS) r-band \citep{Gwyn2012}, we show that our new approach outperforms previous CNN architectures.  \textsc{Mummi} is built to handle both the multi-stage problem (i.e. including both pre-mergers and post-mergers of various epochs), and to greatly reduce the false positive rate (compared to a single CNN architecture) thus improving considerably the veto stage.

The present paper is organized as follows: In \S~\ref{sec2-data} we describe the datasets used, including our merger selections in IllustrisTNG100-1, the generation of UNIONS-realistic mock images and the real UNIONS data to which the trained models will be ultimately applied. In \S~\ref{sec3-methods} we discuss our ensemble of models, \textsc{Mummi}, including a brief overview of Vision Transformers, CNNs, and the adopted strategy to combine the models. In \S~\ref{sec4-results} we discuss the performance of our models in the context of the simulations, as well as how these models can decrease the false positive rate when applied to surveys compared to previous literature. Additionally, we apply \textsc{Mummi} to UNIONS r-band images and report a large sample of merger candidates. In \S~\ref{sec5-discussion} we briefly discuss the representations generated by \textsc{Mummi} and how they can be used as a way to bridge the simulations and observations. We finish in \S~\ref{sec6-conclusion} with a summary of our findings. 

We assume the same cosmological model used by IllustrisTNG, which is consistent with the \citet{Planck2018} results that show $\Omega_{\Lambda,0} = 0.6911$, $\Omega_{m,0} =  0.3089$, and $h = 0.6774$. 

\section{Data}\label{sec2-data}

The goal of this paper is to showcase \textsc{Mummi}, a deep learning method that produces pure merger classifications on UNIONS. To train and evaluate its performance, we created a mock image dataset of 6.4 million galaxy images using the high-resolution large box cosmological simulation IllustrisTNG100-1, targeting the image quality and properties of UNIONS r-band. By using simulations instead of observations, we confidently determine if a particular galaxy is a merging system\footnote{Although infrequent, some numerical issues can produce spurious merging systems in simulations. Some of these issues are: numerical stripping, problems with the subhalo finder and lack of information beyond $z\sim0$ \citep[see ][]{Rodriguez-Gomes2015}.}. Additionally, we can identify its merger stage (pair or post-merger) and the associated timescales with each merger event. 

%Using the simulated data, we create a mock image dataset that post-processes the simulated galaxies to resemble observations by the CFHT (Canada France Hawaii Telescope). This mimics the latest data release from CFIS, utilizing the \textsc{RealSim} code with additional custom steps.
We discuss UNIONS and how it was used in the mock generation pipeline in \S~\ref{subsec:CFIS}, as well as our target dataset where we apply \textsc{Mummi} to find merger candidates. A brief description of the IllustrisTNG simulations, our merger selections and the control matching of our training sample is given in \S~\ref{subsec:TNG-sample}, as well as the pipeline itself with RealSimCFIS in \S~\ref{subsec:realsim}. Finally, we describe the evaluation dataset used as in a \textit{mock survey} in \S~\ref{subsec:mocksurvey}.

\subsection{The UNIONS Canada France Imaging Survey}\label{subsec:CFIS}

Throughout this work, we create bespoke mock observations of IllustrisTNG galaxies to simulate how they might appear in r-band of the UNIONS consortium of wide-field imaging surveys in the northern hemisphere. The r-band imaging was taken at the 3.6-metre Canada France Hawaii Telescope (CFHT) on Maunakea, mapping 4,861 $deg^2$ of the northern sky in $u$ and $r$ filters reaching a \textbf{5-sigma} depth in $r$-band of $28.4$ mag arcsec$^{-2}$.

The r-band observing pattern uses three single-exposure visits with field-of-view offsets in between for optimal astrometric and photometric calibration with respect to observing conditions. This also ensures that the entire survey footprint, including areas in the "chip gaps" between MegaCam's multiple CCDs for a given exposure, will be visited for at least two exposures. After raw images are collected by CFHT, they are detrended (i.e., the bias is removed and the images are flat-fielded using night sky flats) with the software package MegaPipe \citep{Gwyn2008, Gwyn2019}. The images are next astrometrically calibrated using Gaia data release 2 \citep{GaiaCollaboration2018} as a reference frame. Pan-STARRS 3$\pi$ $r$-band photometry \citep{Chambers2016} is used to generate a run-by-run differential calibration across the MegaCam mosaic, and an image-by-image absolute calibration. Finally, the individual images are stacked onto an evenly spaced grid of 0.5-degree-square tiles using Pan-STARRS PS1 stars as in-field standards for photometric calibration. The resulting $r$-band images that will be used in our work have a typical 5-$\sigma$ point-source depth of 24.85 mag, ~0.6-arcsecond seeing, and a pixel scale of 0.187 arcseconds. 

Ultimately, we want to apply the models trained with the UNIONS r-band mock images on the real UNIONS r-band observations. Hence, we use the overlapping sources between UNIONS Data Release 5 (DR5) and the spectroscopic data from the Sloan Digital Sky Survey (SDSS) Data Release 7 (DR7) \citep{Abazajian2009} to probe galaxies from  $z=0.01$ to 0.3. The lower limit of the redshift range ensures that we include only galaxies fitting within our field of view, while the upper limit is imposed due to resolution limitations and to minimize the inclusion of unresolved point-like sources at higher redshifts, such as quasars, in the final sample. The resulting UNIONS r-band DR5 -SDSS DR7 overlap results in a sample of 235,354 real sources with spectroscopic redshifts.

While focusing only on spectroscopic redshifts from SDSS DR7, we limit our sample to the shallower depth of the SDSS survey. However, we still detect faint features of these galaxies at the UNIONS r-band depth, which reveals rich interaction signatures that could not be previously detected. We still lack a robust photometric redshift catalog for UNIONS, but we plan to expand this analysis to fainter sources that are undetected in SDSS once it is available. 

%, derived stellar masses and star formation rates.

\subsection{IllustrisTNG 100-1}\label{subsec:TNG-sample}

IllustrisTNG is a collection of cosmological, gravomagneto-hydrodynamical simulations with a range of particle resolutions. These simulations are conducted in three comoving boxes of $\sim50$, $\sim100$, and $\sim300 \rm \ Mpc~h^{-1}$ length size, known respectively as TNG50, TNG100, and TNG300 \citep{Marinacci2018, Naiman2018, Nelson2018, Nelson2019, Pillepich2018, Springel2018}. We use the TNG100-1 simulation, chosen for its optimal balance between resolution and volume. 

TNG100-1 has been used extensively in studies of galaxy structure and morphology, including the comparison between simulations and observations \citep{Huertas-Company2019, Blumenthal2020}, the importance of interacting galaxies to galaxy properties \citep{Patton2020, Hani2020, Salvatore2021, Quai2023, Byrne-Mamahit2023, Byrne-MamahitB}, and in conjunction with deep learning techniques \citep{Wang2020, Ferreira2020, Bottrell2022b, Ferreira2022}. Specifically, \citet{Zanisi2021} and 
\citet{ERGOML} use unsupervised learning to compare features from the simulation to observations, demonstrating that TNG100-1 effectively reproduces the general morphological appearance of large, massive and extended galaxies. However, discrepancies are expected at small scales due to limited star formation treatment and resolution. For example, \citet{Flores-Freitas2022} and \citet{Whitney2021} show that the inner region of simulated galaxies do not get as compact and concentrated as real ones. These limitations should bare minimum impact in the work presented here since we are focused on the morphological signatures of merging galaxies which are not impacted by these issues.

In the subsequent subsections, we will detail our selection approach, outlining how we extract information from merger trees to determine whether a galaxy has undergone a merger. Additionally, we describe our methodology for matching these galaxies with suitable non-interacting control samples

%While there are some deviations in the small-scale structure of highly concentrated spheroidal systems \citep{Whitney2021}, this is a minor issue in our analysis since they only take up a small fraction of our sample. In addition, our galaxies are resolution limited at the current redshift of interest, meaning that tiny details of structure are not relevant in this analysis.

\subsubsection{Merger Sample}

The merger selection used here follows the merger-tree search method presented in \citet{Byrne-MamahitB}. The goal is to generate a sample that is broadly representative of morphologies encountered in UNIONS in the redshift range $z=0$ to $0.3$, where most of the target galaxies are. Starting from the pool of all available galaxies in the TNG100-1, we limit our selections to galaxies in the redshift range $z=0$ to $1$. This selection is driven by two considerations: firstly, to acquire the largest possible sample of galaxies without straying from the target redshift range; secondly, at redshifts beyond $z=1$, the merger fraction and rate increase substantially \citep{Duncan2019, Whitney2021}. Moreover, beyond $z=1$, galaxies are statistically and intrinsically different from their lower redshift counterparts \citep{Kartaltepe2015, Whitney2021, Ferreira2023}. These issues not only increases the numerical challenges for the subhalo identification tool, \textsc{SubFind}, but also impacts the accuracy of stellar mass ratios  \citep{Rodriguez-Gomes2015, Byrne-MamahitB}.

To ensure that the subhalos selected are true galaxies and not the result of spurious halo fragmentation episodes or that they underwent significant numerical stripping, we select only those subhalos with \texttt{SubhaloFlag=True}. This flag ensures that each subhalo has a cosmological origin and has a realistic dark matter to total mass ratio. Furthermore, we use a stellar mass cut of $M_* > 10^{10} M_\odot$. This constraint ensures that each galaxy in our sample contains at least $\sim 7,000$ stellar particles, a prerequisite for reproducing realistic visual morphologies in mock imaging \citep{Bottrell2017b}. In total, these selection criteria yields a master sample of 301,891 galaxies, corresponding to approximately $\sim 6,000$ galaxies per snapshot for $50$ snapshots.

Within this master sample, we follow a similar approach to that of \citet{Patton2020}, where galaxy pairs are identified by computing the 3D separation of galaxies to their closest companion, selecting only cases with stellar mass of at least 1:10, that is, the stellar mass ratio, $\mu$, has to obey $\mu \geq 0.1$. However, measuring $\mu$ accurately can be challenging due to numerical issues, especially when subhalos are in close proximity and their stellar and dark matter particles overlap \citep{Rodriguez-Gomes2015, Patton2020, Byrne-MamahitB}. To mitigate this, $\mu$ is always calculated based on each galaxy's maximum mass over the previous 500 Myr. Of the 301,891 galaxies in the TNG sample, 272,307 have a companion with a stellar mass ratio of $\mu \geq 0.1$ within 2 Mpc. However, to establish if these will lead to a merging event, one needs to search the merger trees created by the \textsc{SUBLINK} algorithm \citep[see][]{Rodriguez-Gomes2015}. Merger trees trace the hierarchical assembly of each subhalo at $z=0$ back to the preceding simulation snapshots.  Each galaxy is associated with its progenitors and/or descendant galaxies. Our merger selection relies on these associations, enabling us to determine the timescale of a merger event with an accuracy of approximately $\sim$160 Myr, corresponding to the average snapshot time resolution between $z=1$ and $z=0$. 

Using the merger trees, we apply the following additional condition. For post-mergers, to further mitigate issues of mass stripping and exchange, we use the methodology outlined in \cite{Byrne-MamahitB}. Instead of requiring that the masses of interacting galaxies are measured by their maximum past mass \citep{Rodriguez-Gomes2015}, we require their stellar masses to be measured while the merging galaxies are at least 50 kpc apart. In this way, the stellar mass ratio is not calculated after a significant (numerical or physical) mass exchange between the merging galaxies,  avoiding cases where the maximum past mass is found in the distant past of a galaxy merger history, thus not being a representative of the mass during the merging event.

Finally, having aggregated information from the merger trees, we proceed to define our merger sample, encompassing both pre-mergers and post-mergers. Recognizing that some galaxies in pre-coalescence configurations may also qualify as post-mergers, we establish a class priority in our selection criteria. Accordingly, if a pre-merging pair includes a galaxy that could be categorized as a post-merger, it is still labeled as a pre-merging pair. This is done since the pair-phase interaction dominates the determination of a galaxy's morphology. Following this, we identify the post-mergers, and the remaining galaxies generate the pool available for control matching.

It is important to note that there is imperfect overlap between our pre-merger and pair samples: while all pre-mergers qualify as pairs, not all pairs are necessarily pre-mergers. In some cases, this is a result of the finite time of the simulation, since pairs may merge after $z=0$. In other cases, the pairs will not merge despite interacting at close separations for billions of years \citep{Patton2024}. For the purposes of this research, it is crucial to understand that the distinction between pairs and pre-mergers does \textit{not} guarantee non-merging status for our galaxy pairs. Ultimately, both the pre-merger stage and pairs are incorporated into our framework, primarily aimed at reducing the occurrence of false positive identifications of post-mergers, and are not the primary target of our science.

In TNG100-1, the high end of the stellar mass function is dominated by galaxies that had a recent merger or are going to merge in a short timescale. To avoid control matching issues due to the lack of controls compared to mergers in the high mass end, we further apply an upper stellar mass cut of $M_* \leq 10^{11} M_\odot$ to our master sample, reducing the total number of galaxies from 301,891 to 238,674. From this available pool, we select our pre-mergers in two ways. Firstly, we select all galaxy pairs with $\mu > 0.1$ and that merge within the next $1.7\rm~Gyr$, corresponding to ten simulation snapshots.  Secondly, we select pairs that could potentially merge beyond $z\sim0$. This includes cases where no merging event was found, but where the distance to the nearest companion ($r_1$) with $M_* > 10^{9} M_\odot$ is under $50~\rm~kpc$ of the source of interest. Due to their proximity, these are candidates with a high likelihood of merging in the near future, beyond the end of the simulation \cite[e.g. see  Fig. 15 in][]{Patton2024}.

This yields a total of 17,894 pre-merger galaxies. Post-mergers are subsequently selected from the remaining pool of 220,780 galaxies, excluding the 17,894 pre-mergers. We select all cases where a merging event of at least $\mu \geq 0.1$ happened in the past $1.7\rm~Gyr$, and no companions are found within the field of view ($r_1 > 50~\rm~kpc$), resulting in 21,485 post-mergers. 

It is important to highlight that the timescales associated with the merging events in this study are significantly longer than previous works. The use of such a long timescale ($\sim3.4~\rm Gyr$, including pre and post merger stages) is motivated by the findings from \citet{Bickley2021} and \citet{Ferreira2022},  which indicate a higher occurrence of false positives amongst galaxies with slightly higher timescales than the defined window. Incorporating pre-mergers well before coalescence and post-mergers well after coalescence allows us to investigate the connection between observable morphology and temporal proximity of the merger event. To have the future capacity to assess our results as a function of time, we subdivide the post-mergers and pre-mergers in four timescale bins, namely $T~\leq~0.25$ Gyr, $0.25 < T \leq 0.75$ Gyr, $0.75 < T \leq 1.25$ Gyr, and $1.25 < T < 1.75$ Gyr. Pre-mergers with potential merging events beyond $z=0$ are assigned no timescale and excluded from any timescale comparison.

Ultimately, our merger sample includes 39,379 merging galaxies, 17,894 of these are pre-mergers while 21,485 are post-mergers, $\pm$1.7 Gyr before and after coalescence, respectively. 

\subsubsection{Control Matching}

To construct a balanced dataset to train \textsc{Mummi}, we match galaxies that do not fit our merging and interaction criteria to each of our 39,379 mergers. After excluding all the selected mergers from the remaining pool, we identify two types of non-interacting galaxies. The first type, true non-mergers, include galaxies whose most recent merging event occurred at least 1.7 Gyr ago or beyond. The second type, potential non-mergers, includes galaxies with no merging event found within 1.7 Gyr, but that could potentially merge in the future (after the end of the simulation). For these galaxies, we impose a minimum separation from the nearest neighbour of $r~>~50~ \rm kpc$, ensuring that the pairs are sufficiently distant to be unlikely to merge within the specified timeframe, and are out of field of view of the images. For example, \citet{Patton2024} shows that only $30\%$ of pairs at this distance merge within $1$ Gyr. This criterion is essential because we lack future merger information for low redshift systems, and excluding this population would bias our control sample towards higher redshifts. Thus,  this pool of controls yields 168,549 galaxies.

From the control pool, we match each galaxy merger based on stellar mass ($M_*$), redshift ($z$) and gas fraction ($f_\text{gas}$). This matching ensures that the merger and control samples have similar physical properties that could lead to morphological features and are at the same cosmic time. We define $f_\text{gas}$ as
\begin{equation}
    f_{gas} = \frac{M_{gas}}{M_{gas}+M_*},
\end{equation}
with $M_{gas}$ and $M_*$ extracted from the subfield for gas and stellar particles in\texttt{SubhaloMassInRadType} respectively, both from Illustris metadata.
We impose strict matching limits of $\pm 0.05~\rm dex$, identical redshift/snapshot, and $\pm 0.05$ for $f_{gas}$, respectively. In an iterative process, for each galaxy merger, we apply these limits to the control pool and then minimize the distance in the $M_*-f_{gas}$ plane. This approach enables us to select the most suitable control among the control pool. The candidate is removed from the pool of controls so that each merger is matched to a unique control galaxy, and the process continues until all galaxy mergers have a matched non-interacting galaxy.

In summary, each selected merger is matched to a non-merging control galaxy on stellar mass ($M_*$), redshift ($z$) and gas fraction ($f_{gas}$), resulting in $39,379$ mergers and $39,379$ controls.

\subsubsection{\textsc{Mummi} master sample}\label{subsec:mastersample}

After control matching, the 39,379 controls are grouped together with the galaxy mergers to produce a balanced dataset of 78,758 individual galaxies from TNG100-1. As we will describe further in \S~\ref{subsec:realsim}, we generate 80 different mock realizations for each of these galaxies, consisting of 4 different viewing angles and 20 different redshifts, ranging from $z=0.015$
to $0.256$. This process results in 6,300,640 mock images, which are used to train and validate \textsc{Mummi}. However, to prevent information leakage between the training and validation sets, which could occur if different realizations of the same galaxy were distributed across both sets, we perform the train/test split based on the unique ID of each of the 78,758 galaxies. This means that if a galaxy is in the training set, all its realizations are also in the training set. However, we do not separate them at the merger tree level. The same galaxy in different snapshots can be found split between train and test. Their morphologies can change dramatically from snapshot to snapshot due to the coarse time resolution of the simulation, and thus do not constitute an overfitting problem for our models. A breakdown of our sample selection, control matching and training split is shown in Table~\ref{tab:samplestats}.

\begin{table}
\caption{Summary breakdown of our TNG100-1 sample.}
\label{tab:samplestats}
\begin{tabular}{lcccc}
\hline
Set & Pre-Merger  & Post-Merger & Non-Merger & Total \\
\hline
Training & 14,280 & 16,919 & 31,240 & 62,430 \\
Validation & 4,212 & 3,612 & 7,786 & 15,610 \\
Mock Survey & 28,796 & 25,487 & 248,309 & 302,592 \\
\hline
\end{tabular}
\textbf{Note. } The numbers in this table represent the sample before each galaxy was
post-processed with \textsc{RealSim}. For training and validation sample, each galaxy has 80 different realization (4 orientations, 20 redshifts). See \S~\ref{subsec:realsim} for details. A breakdown of our selection, control matching and sample split is described in \S~\ref{subsec:mastersample}.
\end{table}
 
\begin{figure*}
    \centering
    \includegraphics[width=1.0\textwidth]{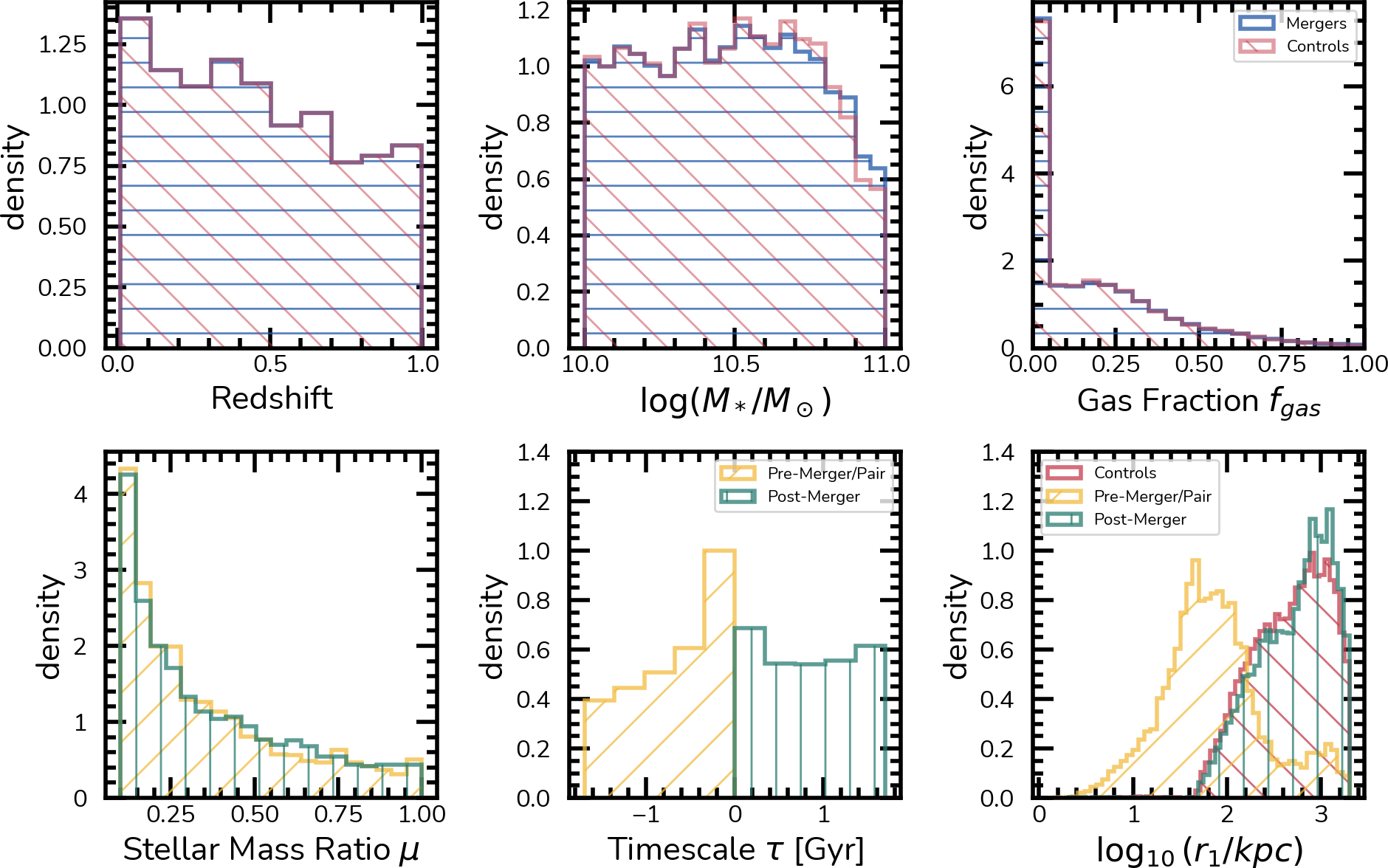}
    \caption{Summary of the selected sample of merging galaxies and its matched controls. The top row displays the distribution of redshifts, stellar masses and gas fractions for the mergers and controls. The matching scheme results in similar distributions for both samples, with minor deviations at the high stellar mass end, remaining within $0.05~\rm dex$ limit imposed. The bottom row displays the overall properties of the pre-mergers and post-mergers in the sample. It shows the stellar mass ratio, $\mu$, the timescales, $\tau$, and the separation to the nearest neighbour, $r_1$. For the mass ratios in the left panel, the small peak at $\mu \sim 0$ corresponds to the pre-mergers without available information, the ones likely to merge beyond $z\sim0$. For the neighbour separations, we also display the distribution of $r_1$ for the controls, illustrating they have very distinct configurations. By definition, the post-mergers and controls lack a companion within $50~\rm kpc$, a criterion inherent to the pair and pre-merger selection. However, there is a small overlap between pre-mergers and the rest of the population for companions at separations larger than $50~\rm kpc$.}
    \label{fig:sample_stats}
\end{figure*}

We show the master sample statistics in Fig.~\ref{fig:sample_stats}. The top row displays the agreement between our merger and control samples for redshift, stellar mass $M_*$ and gas fraction $f_{gas}$. The mergers and their controls are exactly matched in redshift by design, clearly shown in the top left histogram. The stellar masses ($M_*$) are also in good agreement within $0.05~\rm dex$, with a small discrepancy at high stellar masses. The gas fraction $f_{gas}$ is matched within $0.05$ as well. The bottom row shows statistics for the mergers, with stellar mass ratios ($\mu$) on the left, time until/since merger in the middle ($\tau$), and distance to the nearest neighbour with $M_* > 10^9 M_\odot$ ($r_1$) in the right. Here $r_1$ highlights the different environments that pairs/pre-mergers and post-mergers/controls reside in. By design post-mergers and controls lack a companion within $50~\rm kpc$, while pre-mergers/pairs can have close or distant companions.

\subsection{Realistic Mock Pipeline}\label{subsec:realsim}

In this work, we follow the mock observation methodology described in \citet{Bickley2021}, and use only the $r$-band imaging in creating mock galaxy images. We focus on single-band optical imaging to avoid biases that can be introduced by colour images, as well as restricting the wavelength to a regime that is not impacted by extreme dust attenuation \citep{Bottrell2019, SMethurst2002}.  The mock observation pipeline, \textsc{RealSimCFIS}, is based on the \textsc{RealSim}\footnote{Availale at \url{https://github.com/cbottrell/RealSim}} software first used to make SDSS mock observations in \citep{Bottrell2019}, with several modifications. We briefly describe these changes in the following subsections.

\subsubsection{Stellar mass maps}

The inputs for our mock observation pipeline are stellar mass maps, rendered at four camera angles (located at the vertices of a tetrahedron aligned with the simulation box) for each galaxy in the simulation. The stellar mass maps are square, have a physical scale of $100$ kpc on a side, and are rendered with arbitrarily high resolution (2048$\times$2048 pixels) so that no information is lost at this stage. The stellar mass maps are then converted to luminosity maps using \citet{Nelson2019} photometric tables. The light is distributed proportionally to the stellar mass, such that the total brightness matches the absolute $r$-band magnitude.

\subsubsection{Redshift dimming}

In order to represent the observational redshift range of galaxies in our UNIONS-SDSS overlap observation dataset, we define a set of 20 redshifts, spanning $0 < z \leq0.3$. The redshift mesh is not uniformly spaced, rather it is chosen so that each redshift represents an equal number of galaxies in the catalog shared by SDSS DR7 and UNIONS DR5. For each camera angle in each galaxy, we generate all 20 redshift realizations in the set. Each image is realistically dimmed by a factor of (1 + $z$)$^{-5}$, acountting for the usual $(1+z)^{-4}$ cosmological dimming factor \citep{Tolman1930, Disney2012}, plus a bandpass shift correction for the broadening of rest-frame spectral energy density in an observer-frame bandpass of $(1+z)^{-1}$.

\subsubsection{Rebinning and Point-spread function (PSF)}

We next re-bin the light from the redshift-dimmed image to CFHT's MegaCam actual CCD pixel scale of 0.187 arcsec pix$^{-1}$, conserving the total flux based on the angular size of a 100 kpc field of view at the chosen target redshift. Additionally, we choose a UNIONS r-band tile at random, and from the real galaxies therein we select one at random and sample a PSF full-width at half-max (FWHM) corresponding to it, convolving the synthetic galaxy image with a gaussian kernel of that size. 

\subsubsection{UNIONS r-band skies}

The final step in the mock observation pipeline is to select an appropriately sized region of the sky from UNIONS r-band images, and to add it to the image as the sky background. We pre-select at random 100 r-band tiles, and identify a real "proxy" galaxy from the catalog of SDSS DR7 galaxies within UNIONS, and use that galaxy's tile as the sky background for the current mock observation. An 11-arcminute-square cutout is generated at the coordinates of the proxy galaxy, and Source Extractor \citep{Bertin1996} is used to identify a location on the cutout where the centre of the image is not occupied by a source. This approach allows for realistic overlap and crowding effects in the mock images. The UNIONS sky at the chosen location, which is by definition already at the CFHT CCD scale, is added to the mock observation. Any image artifacts are preserved in the process, including saturated stars, missing sky coverage, CCD defects, and zero-flux artifacts. In the rare case that the final image is dominated (more than half) by zero-flux artifacts, the synthetic observation is discarded and re-attempted.

\subsubsection{Final Data Products}

\begin{figure*}
    \centering
    \includegraphics[width=1\textwidth]{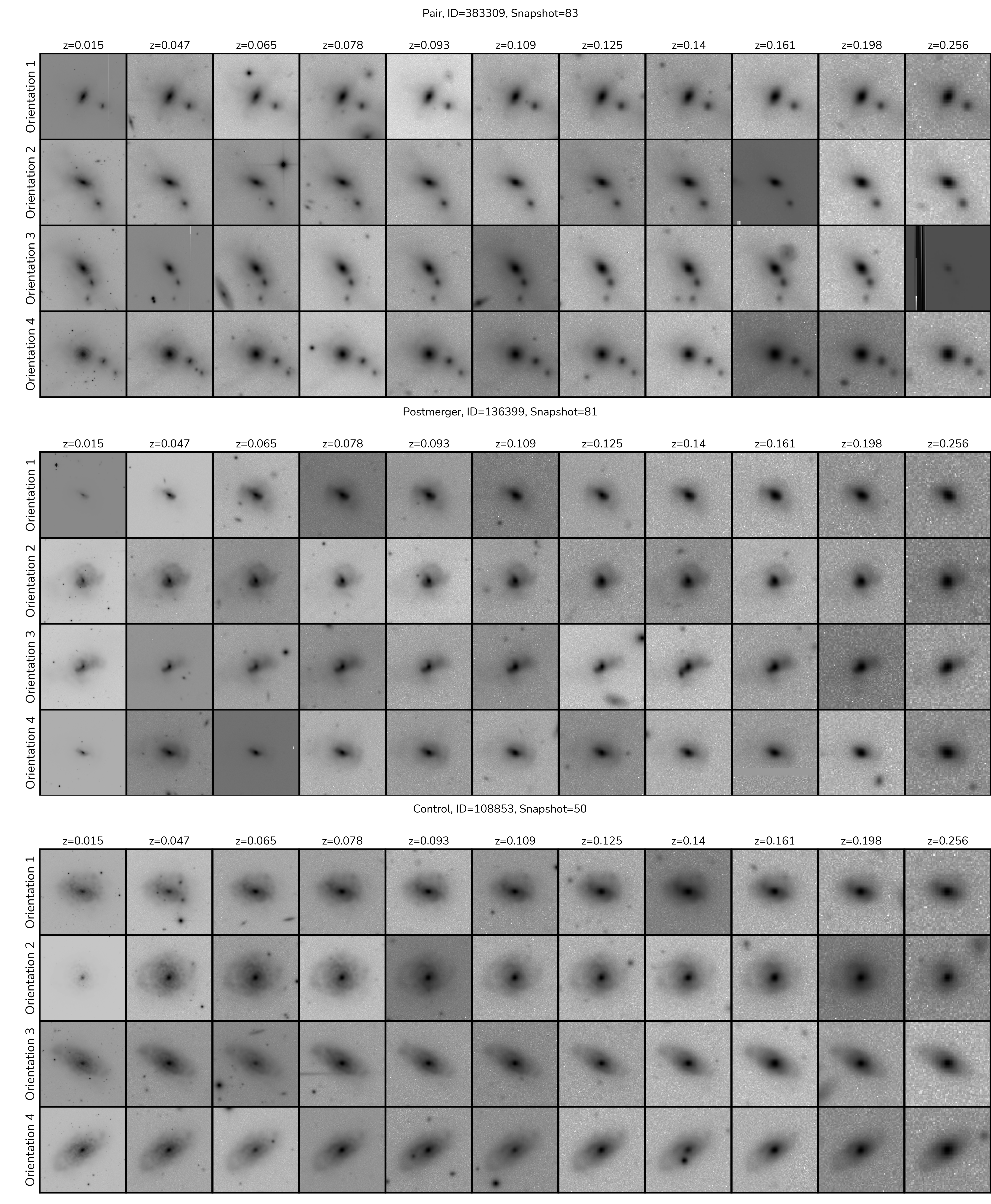}
    \caption{UNIONS r-band mock observations from IllustrisTNG galaxies. Shown are three examples of galaxies in our \textsc{Cfis-Tng} mock dataset, one for each type in the sample, pre-merger (top, ID=383309), post-merger (middle, ID=136399) and control (bottom, ID=108853). Each galaxy is post-processed in 80 different realizations, in 4 orientations and 20 different redshifts. Here we show a subset of 44 of them. A new region of the UNIONS r-band tileset is used for the realistic background of each realization, introducing variability into the mocks, including projection effects, stars, and artefacts similar to what is found in the UNIONS r-band reductions. The redshift range from $z=0.015$ to 0.256 also give us variable depth due to cosmological dimming effects.}
    \label{fig:mock_images}
\end{figure*}

We apply the process outlined in \S~\ref{subsec:realsim} for all galaxies in our selection (\S~\ref{subsec:mastersample}), resulting in 6.3 million mock images with UNIONS-like features. For training our models, we use all mocks associated with the galaxies in our selections as described in \S~\ref{subsec:TNG-sample}. In Figure~\ref{fig:mock_images} we show three examples of galaxies in our selections and all its different mock realizations created by this pipeline. We show a pair in the top set of panels, a post-merger in the middle set of panels and control galaxy in the bottom set of panels. Each row corresponds to a different orientation while each column showcases a different redshift. The mock observations for a given galaxy are in general similar due to the central source, but the process introduces realistic features from UNIONS, such as stars, artefacts, hot pixels, and projected companions.

\subsection{Mock Survey}\label{subsec:mocksurvey}

For experimental validation, we also use the mock survey originally developed for \citet{Bickley2021}, which mimics a realistic distribution of mergers to non-merging galaxies, not following the balanced design of our training dataset. The survey contains a single synthetic observation of every galaxy in TNG100-1 with a stellar mass of M$\star$ $\geq$ 10$^{10}$ M$_{\odot}$ from a fifth camera angle, at a vertex in the first octant of a cube with the galaxy at its centre. Since galaxies are randomly oriented with respect to the simulation box, this camera angle is consistent with a random set of new orientations. New mock observation redshift values are selected at random for each object from the distribution of SDSS DR7 galaxies in UNIONS. The  mock survey contains one image each for 303,110 TNG100-1 galaxies.

\section{Methods} \label{sec3-methods}

Merger identification is a challenging task for automated tools. The rarity of these events in the local universe renders merger identification a highly imbalanced task, making up less than $5\%$ of the whole population of galaxies \citep{Casteels2014}. Therefore, classification tools must achieve high purity as a way to mitigate the rate of false positives that can plague merging selections. Some previous works have therefore used a CNN to do a first pass on large datasets, followed by visual inspection to remove remaining contamination  \citep{Bickley2021, Bottrell2022, PearsonEAS}.  In this work, we attempt to improve the automated (i.e. first step) of this process further, increasing the purity and thus reducing the amount of human intervention required.  Specifically, we propose the use of a simulation-based, hierarchical hybrid deep learning framework: simulation-based as it is trained on simulated galaxies only; hierarchical for its multiple-step process, identifying mergers first and their stage later; and hybrid for mixing two types of distinct deep learning architectures, CNNs and vision transformers (ViTs), respectively. 

We briefly discuss the deep learning architectures adopted for this work in \S~\ref{subsec3:models}. Subsequently, we present how they are used together and the general concept of the framework in \S~\ref{subsec3:multi-step},  while in \S~\ref{subsec3:ensemble} we delve into our ensemble of models and how they are integrated. Finally, we detail the training setup adopted, including hardware, training strategy, hyper-parameters, and data augmentations in Appendix~\ref{subsec3:training}. 

\subsection{Deep Learning Model Architectures} \label{subsec3:models}

\subsubsection{Convolutional Neural Networks}

Classification tasks in astronomy based on deep learning supervised models have employed CNNs to great success as an end-to-end method \citep{Pearson2019, Wang2020, Alexandra2020, Alexandra2021, Bickley2021, Walmsley2022, Bickley2022, Pearson2023, Bickley2023}, and have become standard practice in the field. These models can learn from pre-labeled data using the gradient descent optimization algorithm, which interactively adjusts learnable weights to minimize a cost function that compares models' predictions to actual labels \citep{SGDs}. In simpler terms, CNNs work by scanning images with filters that identify patterns and features, such as edges or textures, which are crucial for distinguishing between different types of galaxies. These patterns are extracted from images using learnable convolution kernels. A plethora of different types of CNNs is available, each with their own benefits and drawbacks. For galaxy merger classification, multiple works employed simple AlexNet-like architectures \citep{Ferreira2020, Bickley2021}, ResNets \citep{Bottrell2022b} and more recently EfficientNets \citep{Walmsley2022, Walmsley2023}.

For this work, we adopt the EfficientNet architecture \citep{EfficientNetOriginal}, which aggregates many improvements over the original AlexNet \citep{AlexNet} implementations, such as skip connections (ResNets) and depth-wise convolutions. Our choice is driven by the fact that, with datasets spanning multi-million images as in our case, EfficientNets demonstrate superior performance with lower computational demand compared to other architectures of the same size. Smaller AlexNets, like those used in \citet{Ferreira2020} and \citet{Bickley2021}, 
do not generalize effectively with such a large datasets, resulting in underperformance by $10\%$ margins. Specifically,  we use the fiducial \texttt{EfficientNetB0} implementation from the \texttt{keras} package, with the standard classification head for binary classification tasks.

\subsubsection{The Shifted-Window Vision Transformer}

A different kind of architecture based on self-attention is gaining traction in machine vision. In contrast to convolutions, self-attention, when applied to images, creates correlations among small patches of the input with itself, as follows

\begin{equation}
    \text{Attention}(Q, K, V) = \text{softmax}\left(\frac{QK^T}{\sqrt{d_k}}\right)V, \\
\end{equation}
where $Q$, $K$ and $V$ represent matrices of weights applied to the inputs, known respectively as the query, the keys, and the values. This process generates attention maps across the entire input, highlighting regions crucial for label prediction, which are preferentially focused on by the adjusted weights during training. 

Deep learning architectures that incorporate this mechanism, known as Transformers, were originally developed for natural language processing tasks like translation \citep{Attention2017}, where they combine attention with multilayer perceptrons. In vision applications, the approach differs slightly. Instead of a stream of textual tokens, images are divided into small patches, which are then encoded in a token-like format. The resulting model, known as a Vision Transformer (ViT), is very similar to its original text-based application \citep{visionTransformer}. Contrary to convolutional approaches that process local pixel neighborhoods, self-attention in ViTs captures long-range dependencies between all parts of the image, irrespective of spatial proximity. In the galaxy merger case, attention maps can, for example, associate tidal features to a particular source in the image, instead of just being extracted as general features. Subsequently, these attention maps are used in the classification task, in a similar manner to the features extracted from convolutions in a CNN.

%The Although CNNs dominate the share of classification applications in astronomy, the Vision Transformer \citep{visionTransformer} (ViT) and its variants -- an attention-based deep learning architecture -- is gaining substantial traction. ViTs use the self-attention paradigm that allows the model to dynamically focus on different parts of the image, considering each patch in the context of others. This mechanism enables it to learn spatial hierarchies and relationships between different parts of the image, which can be crucial for specific contexts. 

%Briefly, the self-attention in a ViT maps an input image into a series of 'patches', which are then combined with learnable weight matrices $Q$ (query), $K$ (key), and $V$ (value), that generate an attention map using a softmax function between the dot-product of these matrices:

%where $d_k$ is the dimension of each patch of the image. This creates a self-correlation map which is weighted by the learnable parameters optmizing the training task. In the galaxy merger case, attention maps can, for example, associate tidal features to a particular source in the image, instead of just being extracted as general features. Unlike convolutional approaches that process local neighborhoods of pixels, the self-attention in ViTs allows for the capturing of long-range dependencies between any parts of the image, regardless of their spatial proximity. These maps are then used in the classification task, in a similar manner to the features extracted from the convolutions in a CNN. 

However, these benefits of ViTs are accompanied by some downsides. ViTs scale quadratically with the number of pixels, making them more computationally expensive to train compared to CNNs. Additionally, symmetries that are naturally captured by convolutions, like translation equivariance, must be learned during training in ViTs. This increases the data requirements for generalization, making them less useful with small datasets. Yet, we argue that certain symmetries inherent of CNNs do not generalize effectively to the astronomical context. Given the large scale of our sample ($\sim 6.3$ million images), our training is not limited by training data. For comparison, our dataset is approximately $\sim160$ times larger in size than the dataset used in \citet{Bickley2021}. Moreover, in galaxy classification, the main source is usually centered in the inputs. The spatial correlation with neighbouring sources is relevant, which is not directly captured by CNNs. This means that merging features anywhere in the images can be weighted by their positioning when extracted by ViTs.

To tackle the computational cost issue, variations of the original ViT implementation change how the attention maps are made \citep{vitsreview}. A popular example is the Shifted-Window Vision Transformer \citep[SwinT, ][]{SwinT2021}, which, rather than generating attention maps for entire images, divides the inputs into smaller, non-overlapping regions and computes attention maps for each. To establish causal connections between these attention windows, their positioning is strategically shifted and cycled through each network block. Additionally, the sizes of the patches are increased between attention blocks, rendering SwinTs hierarchical in terms of resolution scales. For these reasons, we adopt the Swin Transformer as our ViT architecture of choice. Previously, a few other studies used SwinTs applied to galaxy mergers, namely 
\cite{Margalef-Bentabol2024} and \cite{Pearson2024}, but at a smaller scale.

%This is the first ever application of SwinT into the context of galaxy mergers.

\begin{figure*}
    \centering
    \includegraphics[width=1\textwidth]{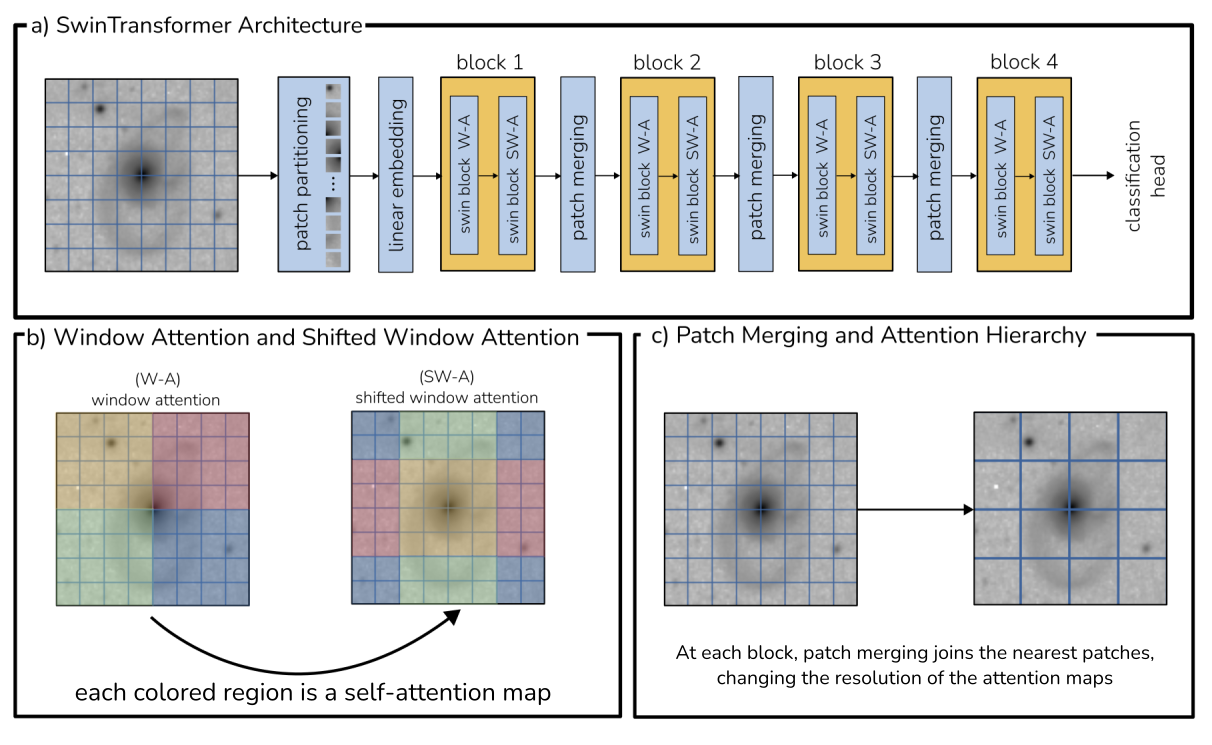}
    \caption{\textbf{SwinTransformer architecture schematic.} We show the general structure of the SwinTransformer architecture we adapted for single-band imaging in panel \textbf{a)}, outlining its main components. Panel \textbf{b)}, exemplifies the window attention and shifted-window attention that is embedded in each of the Swin blocks, while panel \textbf{c)} illustrates the patch merging step between the blocks, where patches are merged together to form larger ones, resulting in lower resolution attention maps.
    } 
    \label{fig:swinTexample}
\end{figure*}

In Figure~\ref{fig:swinTexample} we present a visual summary of the SwinT used in this work. The top panel \textbf{a)} shows the overall sequential architecture, from the input image to the final classification head, and we briefly describe it as follows. First, the input image is divided into small patches in the patch partitioning step. Second, the patches are encoded into one-dimensional vectors of size \texttt{embed\_dim} (a hyper-parameter) through a linear operation. In our case, we use a convolutional layer with \texttt{embed\_dim} filters. Sequentially, these embeddings are then passed through SwinT attention blocks, for a total of four blocks. Each block consists of two steps: the first calculates the window attention, and the second the shifted window attention. This shift slides the attention windows by a few patches to establish the causal connection between the whole input. The bottom-left \textbf{b)} panel exemplifies this, showing the region of each attention window with specific colours, and how these windows slide across the inputs in the second step. Between each block, the outputs from the previous block undergo a patch merging step. This step joins four adjacent patches into one, effectively changing the resolution of the attention maps produced in each block, as exemplified in the bottom-right \textbf{c)} panel. All the attention windows across the architecture contain a fixed number of patches, but the size of these patches, in terms of pixels, varies between blocks. The outputs from the final attention block are then directed to a standard classification head, akin to those used in CNNs. Thus, in the SwinTransformer, the key features are not extracted from convolutions, but are the attention maps themselves.

Here we use \texttt{tfswin}, a \texttt{TensorFlow} SwinTransformer implementation\footnote{Available at \url{https://github.com/shkarupa-alex/tfswin}}. Specifically, we employ the \texttt{SwinTransformerTiny} architecture, but with a custom \texttt{embed\_dim} parameter set to $24$, deviating from the default value of $96$, since our task has considerably fewer classes to be classified than the ImageNet baseline.

\subsubsection{A hybrid approach}\label{subsubsec:hybrid}

While EfficientNets and SwinTransformers can achieve similar classification performances within the same dataset,  Vision Transformers have been shown to outperform CNNs in other downstream tasks, like  image segmentation \citep{mikeSpirals}. Given that our main goal is to classify galaxy mergers, we capitalize on the unique and complimentary information extraction mechanisms of these models. A SwinT model (vision transformer) and an EfficientNet model (CNN) classify the same image based on different information, enabling the leveraging of agreement between these distinct models to enhance the purity of our decision-making process. We explore this in more detail in \S~\ref{sec4-results}. In our framework, each task consistently employs a pair of CNN-SwinT models or an ensemble of such pairs, highlighting the hybrid nature of our approach.

%Bickley et al. (2021,2022,2023) train and use a simple 4-layer-deep CNN for post-merger identification in CFIS. The architecture is loosely based on Alexnet (Krizhevsky et al. 2012), with a number of modifications that were beneficial to the model's performance on the datasets explored in those papers. The architecture is outlined in Table~\ref{arx-table}. Like the other models detailed in this work, the model was trained using augmented images (see Section (\textbf{SECTION}), an Adadelta optimizer (Zeiler 2012) with a learning rate of 0.05, and a training batch size of 32 images.

\subsection{\textsc{Mummi}: A multi-step classification framework}\label{subsec3:multi-step}

\textsc{Mummi} employs a multi-step hierarchical framework for galaxy merger classification, utilizing different models trained on distinct subsets of the merger and control samples. The current version of \textsc{Mummi} has two steps. In the first step, a large ensemble of models classifies galaxies as either mergers or non-mergers, irrespective of their merger stage. The second step employs specialized models to determine the merger stage (i.e. classification as either pre-merging pair or post-merger), assuming that contaminants have been filtered out in the first step. A future third step to rank post-mergers by their timescale will be explored in an upcoming paper.

This hierarchical approach contrasts with conventional methods that simultaneously address multiple tasks, such as both identifying mergers and defining their stages. The separation of tasks in \textsc{MUMMI} addresses distinct class imbalance issues for each task. In particular, merging and non-merging galaxies exhibit a substantial imbalance, while pre-merger and post-merger galaxies are relatively balanced when considered as a subset of mergers alone. Combining these tasks in a single model would exacerbate the imbalance problem and increase the rate of false positives. The use of simulated data with true labels from metadata allows for task-specific training sets and optimization objectives. For instance, the first-step models prioritize purity, while the second-step models may focus on completeness. Additionally, each step can have a different ensemble size (i.e. the number of models used together). Subsequently, tasks that are dependent on merger identification or stage determination can be integrated into this hierarchical framework, with each step acting as a filter for other downstream tasks.

The number of models in each step varies according to the specific task requirements. However, each ensemble step contains an even number of models, comprised of CNN-SwinTransformer pairs (see  \S~\ref{subsubsec:hybrid}). As the first step requires high purity, we use a large ensemble with ten pairs of CNN$+$SwinTransformers (twenty models), while the second step, for being more balanced, has only one model pair. 

%To promote diversity in model training, different subsets of the training data are used for ensembles containing multiple pairs. The architectures employed are detailed in the subsequent subsection.

\subsubsection{Jury-based decision-making}\label{subsec3:ensemble}

In addition to our novel pairwise (CNN+SwinTransformer) multi-step method, another improvement over previous work is our approach to classification. In contrast to the usual approach of using deep learning models, which involves either setting a decision probability threshold or employing Monte-Carlo dropout to artificially produce ensembles from a single model, we base our decision-making on the agreement between different models. While deriving a classification from the number of positive predictions in an ensemble might seem akin to adjusting the probability threshold based on the average of all models, our jury-based approach entails distinct implications. Specifically, the training setup is tailored to maximize the purity of this consensus-driven method. 

This approach is inspired by Condorcet's Jury Theorem \citep{Boland1989}, which posits that for an ensemble of classifiers that are better than random, the collective decision of the ensemble usually surpasses the accuracy of any single model. This is contingent on the classifiers being both independent and diverse. Unfortunately, full independence of the models in the ensemble cannot be guaranteed, as they all share data from the same pool. However, we mitigate this by training each model with different subsets of the full training set. This approach ensures that each model's training history is unique, influenced by both the weight initialization and the specific data encountered during training. The requirement for diversity is, at least in part, fulfilled by our hybrid approach, which utilizes different types of machine vision models.

\begin{figure}
    \centering
    \includegraphics[width=0.48\textwidth]{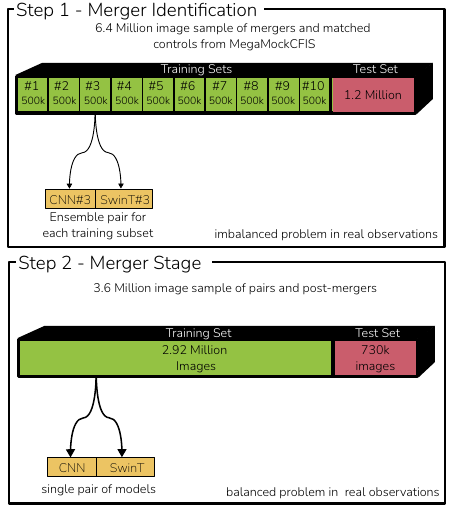}
    \caption{\textbf{Training setup for our multi-step framework}. We divide our merger characterization in two steps, the first one deals with the identification of a merger regardless of its stage amongst all available galaxies (top), while the second step focus only on defining the stage of an already identified mergers (bottom). For the merger identification, we break down our training set in 10 subsets of equal size, and train a pair of CNN-SwinT models for each, then, these 20 models are combined in a voting system for the classification. Given that the identification of mergers in the real universe is very imbalanced, this ensemble of models can be use to increase the purity of the framework. For the second step we assume that the galaxy is already classified as a merger and train only a single pair of CNN-SwinT, as in this case one does need to worry about the imbalance of the identification.}
    \label{fig:ensemble}
\end{figure}

Figure~\ref{fig:ensemble} demonstrates our approach to splitting the training sets for extracting features critical to a jury-based voting system. It shows the division of test and training data, the allocation of dataset subsections for each model pair, and the number of images in each subset. The top panel specifically illustrates this approach for the identification step, while the bottom panel focuses on the process for merger stage classification.

In our framework, we primarily employ this voting system for the merger identification step, where high purity is essential. By doing so, we can select a number of voters out of the ensemble that satisfy our condition. Specifically, in this work, we maximize purity by assuming positive predictions only for unanimous cases, where all models agree, i.e. a galaxy is labelled as a merger only if all 20 models agree. However, we track the votes to facilitate alternative selection procedures in the future. For instance, one can use a binomial distribution with the ensemble probabilities for approaches \textbf{different from the unanimous case}. Just as a positive classification (as a merger) requires agreement by all 20 models in the first step, for the second step, both models in the CNN-SwinT pair must agree to classify a merging galaxy as a post-merger.

\section{Results} \label{sec4-results}

In this section, we report the performance of \textsc{Mummi} in two key areas: merger identification (hereafter \texttt{STEP1},~\ref{subsec:mergeridentification}) and merger stage classification (hereafter \texttt{STEP2}, \S~\ref{subsec:mergerstageclassification}). Moreover, in both cases, we discuss the impact of physical properties, such as stellar mass, environment, and redshift, on the performance of the models. Additionally, in \S~\ref{sec:mockresults}, we apply both steps of \textsc{Mummi} to the mock survey described in \S~\ref{subsec:mocksurvey} to explore our methods in a realistic context. Finally, in \S~\ref{sec:postmergersCFIS}, we present a catalog of unanimously classified post-mergers in the overlap between UNIONS DR5 and SDSS DR7.

All results based on TNG100-1, for which we have ground truths available, are reported using ten-fold cross-validation, with error bars representing $~\pm~1~\sigma$ uncertainties.

\subsection{\texttt{STEP1} Merger Identification}\label{subsec:mergeridentification}

We examine the performance of our ensemble of twenty models for merger identification on the witheld test set of $1,248,800$ images. This comprises $15,610$ unique galaxies, each represented in $80$ individual realizations, combining twenty redshifts with four viewing angles. To assess the performance of \texttt{STEP1} models, we calculate the average probability across all redshifts, yielding 62,440 different classifications. This approach mitigates potential biases emerging from survey artifacts that are introduced during the dataset generation pre-processing step, and allow us to understand any intrinsic uncertainties of the models. Additionally, this also ensues that the reported metrics are not inflated by additional redshift realizations of easy to classify galaxies. For redshift trends, we later explore its impact on the performance by using the different redshift realizations individually. As we include galaxy mergers within long timescales of up to $\pm 1.75$ Gyr around the selected merging events, we first investigate the performance on the whole dataset irrespective of the temporal impact, and later delve into how the performance varies with these timescales. 

\begin{figure}
    \centering
    \includegraphics[width=0.45\textwidth]{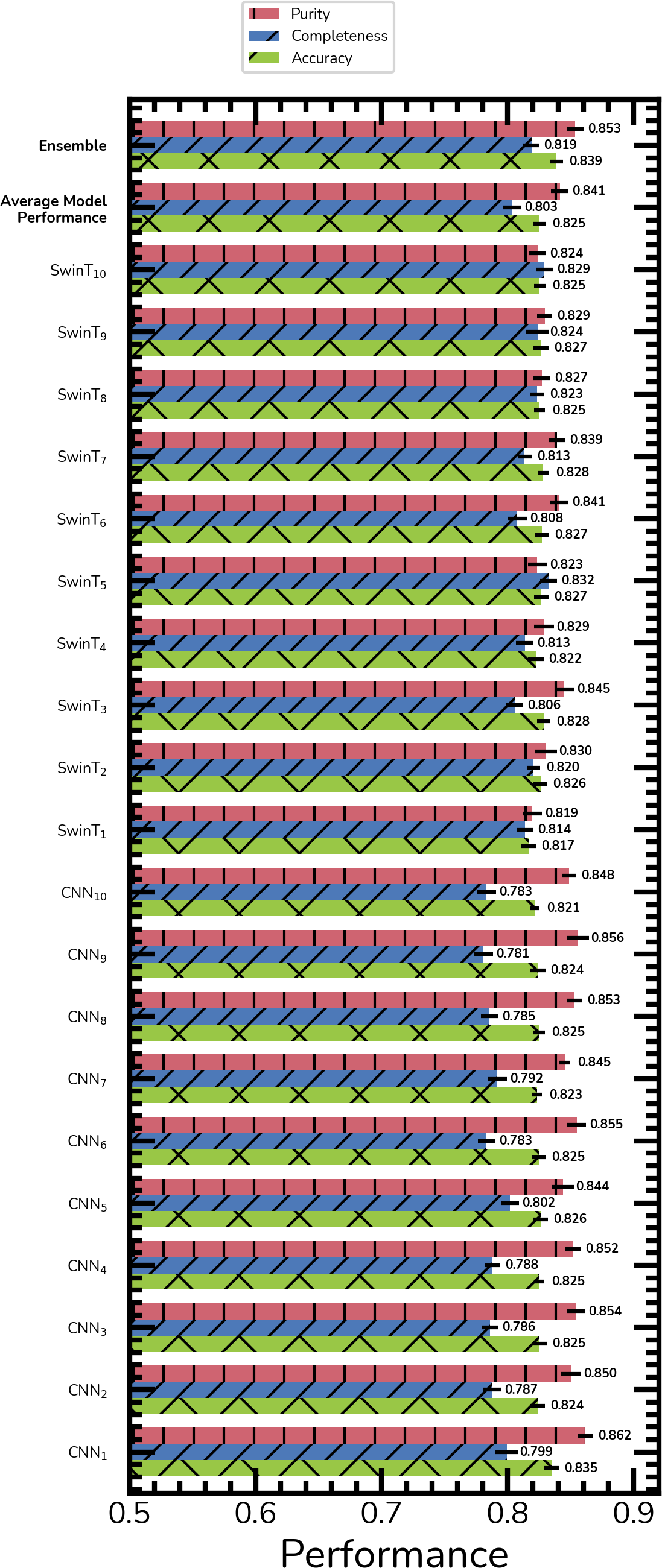}
    \caption{\textbf{Individual Model Performances in Merger Identification}. We show the individual performance of each one of the twenty models in our hybrid CNN-SwinTransformer deep learning ensemble for merger identification, as well as the performance of the ensemble when used together and the average performance for all models. Each bar group represents one of the models trained with the model performance highlighted at the top. The metrics -- purity (red), completeness (blue), accuracy (green) -- are showcased to demonstrate that the resulting ensemble performance is better than the sum of its parts, displaying $\approx 2\%$ improvement over the mean performance of the individual models. The numerical scores are displayed above each bar. Error bars display the $\pm 1~\sigma$ uncertainty as measured from a 10-fold cross-validation in the test set.}
    \label{fig:stage1barplots}
\end{figure}

%First, all individual twenty models in our merger identification ensemble for all the test set, including pairs and post-mergers with up to $1.75$ Gyr timescales and their matched controls. We evaluate the models using a 10-fold cross-validation on the available test set classifications without discriminating the timescales of the merger in the set. Additionally, we evaluate the usage of the ensemble as one entity through a simple majority voting system where a positive merger classification is taken when eleven or more models classify the particular image as a merging system. 
In Figure~\ref{fig:stage1barplots} we report the performance of each individual model in \texttt{STEP1}, separated into CNNs and SwinTransformers. Our performances are tracked through accuracy, purity and completeness statistics \citep{POWERS}. These are shown as bar plot sets, all based on a decision threshold of $P > 0.50$. We include metrics based on the number of True Positives (TP), True Negatives (TN) and False Positives (FP), and False Negatives (FN). We use the purity (in green), defined as 
\begin{equation}
    \rm{Purity} = \frac{TP}{TP+FP},
\end{equation}
the the completeness (blue)
\begin{equation}
    \rm{Completeness} = \frac{TP}{TP+FN},
\end{equation}
and the accuracy (green)
\begin{equation}
    \rm{Accuracy} = \frac{TP+TN}{TP+TN+FP+FN}.
\end{equation}
These indicators help us access both the raw performance, but also the trade-offs and contamination between mergers and non-mergers. Additionally, we include a bar set depicting the average model performance (second from the top) and the performance of whole ensemble with a simple majority voting (top). The ensemble's performance, at $85.3\%\pm0.8\%$ purity, $81.9\%\pm0.9\%$ completeness, and $83.9\%\pm0.5\%$ accuracy, surpasses the average of the individual models' performances, which stands at $84.1\%\pm0.7\%$, $80.3\%\pm1\%$, and $82.5\%\pm0.6\%$, respectively. This indicates an improvement of $1-3\%$, with the ensemble being $1\%$ purer and $2\%$ more complete than the individual models.

Although our models are not completely independent (see \S~\ref{subsec3:ensemble} for a discussion), this shows the Jury theorem in action. Additionally, it is evident that the SwinTransformers generally display a more balanced performance than the CNNs, with similar levels of purity and completeness. The CNNs, on the other hand, tend to skew towards higher purity at expense of completeness. For some CNNs, the purity exceeds that of the ensemble, but this comes with a $5\%$ sacrifice in completeness. While the performance gain of the ensemble might seem marginal compared to the average individual model, it is significant given our goal of applying \textsc{Mummi} to millions of sources in UNIONS DR5. Each percentage point improvement translates to thousands of additional correct classifications, and thus less contamination.

These performance metrics might initially appear to be on par with or lower than those reported in \citet{Bickley2021}, where a simulation-driven CNN used to classify immediate post-mergers ($\tau\approx0$) in UNIONS r-band images achieved an accuracy of $87.6\%$. However, we remind the reader that the results presented here are based on a sample with very long timescales of up to $1.75$ Gyr whereas the \citet{Bickley2021} sample was only for post-mergers that had coalesced within the preceding snapshot. A wider timescale window introduces galaxy mergers that are more challenging to classify. These galaxies often have less distinct merging features, either because they have not been recently close enough for dynamical friction to disrupt their stellar contents, or because such features have diminished over time as the morphology of the descendant galaxy dynamically settles. Our approach contrasts with previous works where very narrow selection windows were adopted, driven by studies on the observability of mergers using non-parametric morphologies \citep{Lotz2008, Whitney2021}.

\begin{figure*}
    \centering
    \includegraphics[width=1\textwidth]{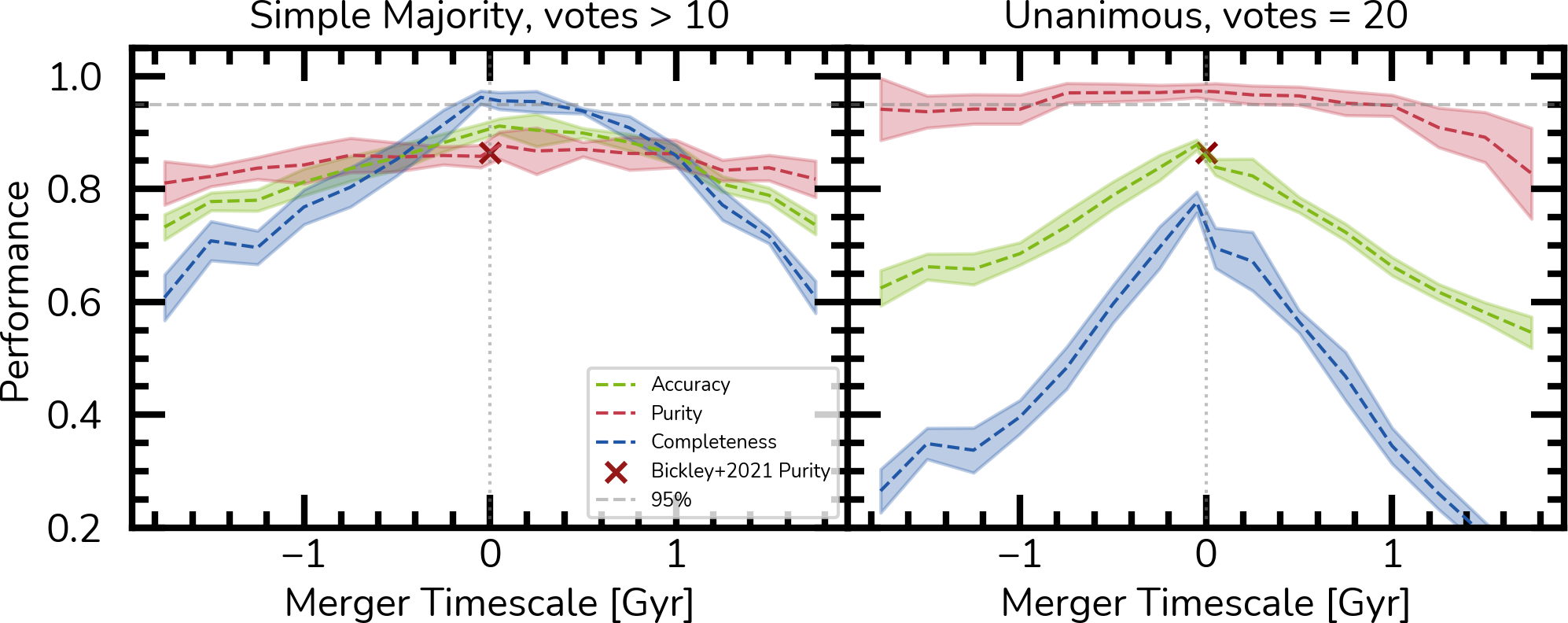}
    \caption{\textbf{Merger identification performance with timescales for simple majority and unanimous votes}. We show how the accuracy (red), purity (green) and completeness (blue) are related to the merger timescales in the test set for a simple majority ($> 10$ votes, left) and for an unanimous selection ($= 20$ votes, right). This showcases the impact of the timescales (time since / to the merger event) on merger observably. Even though the model performance trends negatively with increasing timescales, it can still classify mergers correctly up to $1.75$ Gyr after/before a merging event. By using the unanimous vote, the purity of our model is consistently above $95\%$ for all timescales, peaking at $97.5\%$ at $\tau = 0$ Gyr, demonstrating that by using the unanimous case one can correctly select pure samples of mergers for any timescale, changing only on the completeness of the sample at each timescale. In this way, although our ensemble is less complete at long timescales, its purity is largely unaffected. This behaviour is desirable given that the frequency of mergers in the local universe is very imbalanced when compared to the whole population of galaxies. The horizontal and vertical dashed grey lines show a performance of $95\%$ and a merger timescale $\tau$=0, respectively, to guide the eye.}
    \label{fig:timescaleperformance}
\end{figure*}

To make this point clearer, in Fig.~\ref{fig:timescaleperformance} we show the influence of merger timescales on the performance of \textsc{Mummi}'s \texttt{STEP1}. We display results for two ways of utilizing the ensemble: a simple majority of the models in the ensemble with a standard threshold (left), and establishing an unanimous consensus (right). Fig.~\ref{fig:timescaleperformance}  shows the purity (red), completeness (blue), and accuracy (green), demonstrating their variation with respect to timescales, ranging from the pair phase (indicated by negative timescales) to the post-merger phase (positive timescales). 

Initially, we examine the simple majority case, as shown in the left panel of Fig.~\ref{fig:timescaleperformance}, allowing us to directly compare our results to those of \citet{Bickley2021}, which are represented by a black cross.  At timescales of $\tau \sim 0$ Gyr, our ensemble achieves $86.6\%$ purity, $95.7\%$ completeness, and $90.4\%$ accuracy, respectively. In comparison, \citet{Bickley2021}'s CNN reports $88.5\%$ purity, $86.5\%$ completeness, and $87.6\%$ accuracy.  Although the ensemble exhibits a slightly lower purity, it is consistent within the error bars, and its completeness is $\sim 10\%$ higher. Performance remains comparable to that of \citet{Bickley2021} up to $\pm0.75$ Gyr, but begins to decline beyond $\pm1$ Gyr. Even at longer timescales beyond $\pm1.25$ Gyr, the ensemble maintains consistent purity, albeit with reduced completeness. This indicates that while the ensemble detects fewer mergers at these extended timescales, the associated false positive rate remains comparable.

Using the ensemble with a simple majority can approximate an average probability across the models within it. However, the ensemble can also be utilized in other ways to bias the classifications towards either higher purity or higher completeness. For example, aggregating votes using a binomial distribution is one alternative approach \citep{Walmsley2023}. However, considering the specific case of merging galaxies and their intrinsic imbalance in the real universe, we recommend using the ensemble unanimously. This involves selecting galaxies as mergers only if they are classified as such by every model in the ensemble. 

The right panel of Fig.~\ref{fig:timescaleperformance} shows the results with respect to the timescales for the unanimous case. In contrast to the simple majority case, both accuracy and completeness significantly decrease for longer timescales in the unanimous case. However, the purity remains robustly above $95\%$ for nearly all timescales, except for post-mergers beyond $1.25$ Gyr, peaking at $97.5\%$ at $\tau = 0$. 

The fact that galaxy mergers are rare occurrences in the Universe adds complexity to this picture. To fairly evaluate if \textsc{Mummi} is capable of robustly selecting samples of mergers in the real imbalanced context, we apply the Bayes rule to these metrics \citep{Bayes1763}. We can write the probability of a merger classification from \textsc{Mummi} being correct, $P^{\text{merger}}_{\text{corr}}$, as
\begin{equation}\label{eq:bayes}
P^{\text{merger}}_{\text{corr}} = \frac{\text{Purity} * f_m}{\text{Purity} * f_m + (1-\text{Purity}) * (1-f_m)},
\end{equation}
where $f_m$ is the intrinsic merger fraction. Thus, assuming a $5\%$ merger fraction \citep{Casteels2014} within the redshifts, stellar masses and mass ratios probed in this work, for mergers with short timescales ($\tau < 0.25$ Gyr) \textsc{Mummi} selects samples that are $\sim70\%$ correct. For longer timescales, we find in average that $\textsc{Mummi}$ produces correct classifications at a $\sim58\%$ and $\sim 52\%$ rate, for pairs and post-mergers respectively. In comparison to \citet{Bickley2021}, if we use the their metrics with $f_m = 0.05$, the classifications are only correct at a $\sim30\%$ rate. Even though a large portion of the entire population of mergers is lost with our unanimous approach, the resulting high purity aligns well with the requirements to retrieve a robust sample of mergers amongst the real, imbalanced distribution of galaxies in the Universe. This leads to a low false positive rate, crucial for identifying large samples of merging galaxies in an automated fashion without a visual classification veto (which becomes increasingly untenable for larger samples). Although completeness at timescales of $1.75$ Gyr can drop to approximately $20\%$, this method still effectively identifies galaxies with distinct merging features, thereby minimizing contamination in our samples.

\begin{figure}
    \centering
    \includegraphics[width=0.45\textwidth]{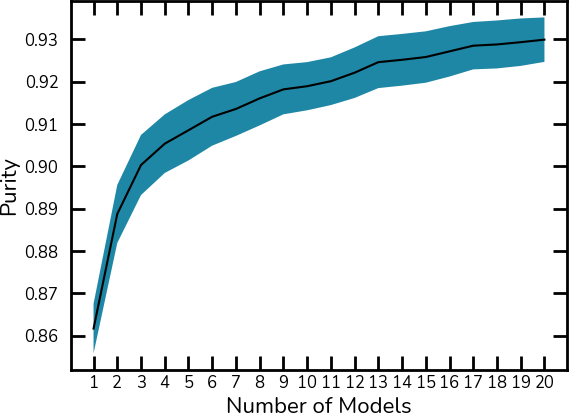}
    \caption{\textbf{Ensemble size vs. Purity.} We show how the ensemble size impacts the purity when classifications need to agree unanimously. Starting from a single model, we re-measure the purity for each new model added to the ensemble. The purity increases rapidly with the initial models, plateauing but slowly increasing. }
    \label{fig:purityvsensemblesize}
\end{figure}

In Fig.~\ref{fig:purityvsensemblesize} we show how the purity increases by increasing the size of the ensemble one by one, ignoring the pair framework we set, where classifications are taken as the unanimous agreement. This illustrates how a large ensemble contributes to increasing the purity in classifications, and serves as a diagnostic tool for the ideal ensemble size to be used. Based on this, the classifications within a balanced dataset presented in this section could be done using only 12 models instead of all 20. 

Ultimately, we recommend the usage of all 20 models if completeness is not an issue, since a $1-2\%$ purity boost from 12 models to 20 represents $10-20\%$ increase in $P^{\text{merger}}_{\text{corr}}$ (Eq.~\ref{eq:bayes}) in the imbalanced case. This makes for a more robust method to be applied at large scale, reducing the necessity for visual inspection. On the other hand, for science cases that result in small samples of mergers where follow up visual classification is viable, the total number of votes used as a threshold can be lowered to increase completeness. However, when applied to large wide-field surveys we recommend the unanimous approach.

%Although demonstrated in a balanced dataset, this approach has the potential to significantly mitigate the false positive rate in real observations.
%\begin{figure}
%    \centering
%    \includegraphics[width=0.47\textwidth]{figs/barplots_models_timescale.png}
%    \caption{Temporal performance of our merger identification compared to \cite{Bickley2021}. We show the purity (green), recall (blue), accuracy (red) and the area under the ROC curve (AUC, yellow) from \cite{Bickley2021}, which is focused only on recent post-mergers, and for five timescale bins as classified by our ensemble, $\tau = 0$ Gyr, $0 < \tau < 0.25$ Gyr, $0.25 < \tau < 0.75$ Gyr, $0.75 < \tau < 1.25$ Gyr, $1.25 < \tau < 1.75$ Gyr, respectively. The merger identification ensemble can correctly classify mergers up to $\approx 0.75$ Gyr with similar performance of that of \cite{Bickley2021}, which is only trained to clssify post-mergers right after the merger event ($\tau = 0$ Gyr). Notably, the model retains a significant accuracy level of $79\%$ for classifying mergers over longer timescales, extending up to $\tau = 1.75$ Gyr.}
%    \label{fig:bartimscales}
%\end{figure}

\subsubsection{Impact of physical properties, environment and redshift}

\begin{figure}
    \centering
    \includegraphics[width=0.45\textwidth]{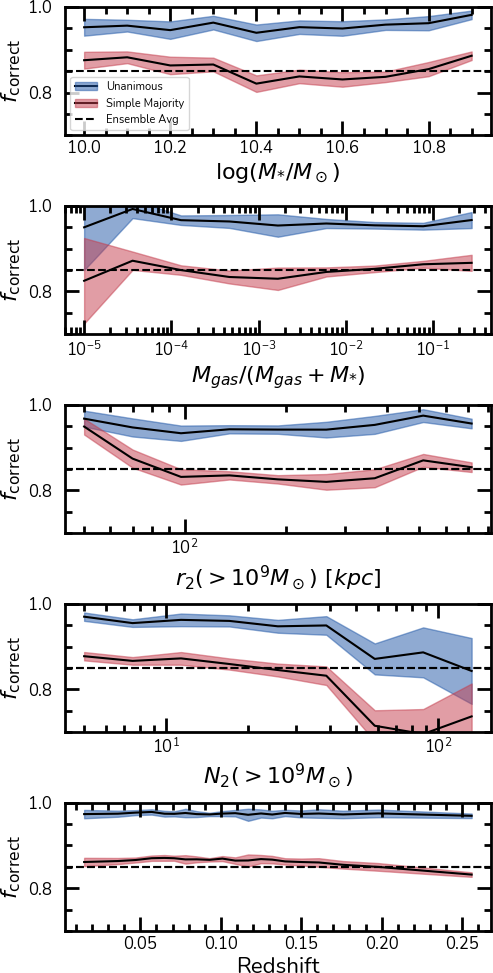}
    \caption{Fraction of correct classifications vs. log stellar mass (top), gas fraction, $f_{gas}$  (second from the top), distance to the second closest neighbour, $r_2$ (middle), number of sources with masses $> 10^9 M_\odot$ within $2~$Mpc, $N_2$ (fourth from the top), and redshift $z$ (bottom). We show the fraction of correct classifications for unanimous voting (blue) and a simple majority (red). Dashed line represents the ensemble average performance. The performance with respect to the stellar masses, gas fractions and redshifts are robust within the domain of our test set, however environment play an important role on the observability of the mergers, with performance decreasing strongly in dense environments. Error bars display the $\pm 1~\sigma$ uncertainty as measured from a 10-fold cross-validation in the test set.}
    \label{fig:environmentstats}
\end{figure}

The individual impact of physical properties, such as stellar mass, and gas fraction, the environment and redshift is compounded and thus hidden when describing the raw performance of \textsc{Mummi}. The models may exhibit biases toward specific domains of these quantities due to slight imbalances in the training datasets, or that particular configurations of mergers are harder to classify. To assess this, Fig.~\ref{fig:environmentstats} presents the fraction of correct classifications ($f_{\rm correct}$), including mergers and non-mergers, in bins of several factors: stellar mass ($M_*$, top), gas fractions ($f_{\rm gas}$, second from the top), their environment, as indicated by the distance to the second closest companion with minimum stellar mass of $10^9 M_\odot$ ($r_2$, middle), the number of neighbours with stellar masses above $10^9 M_\odot$ within $2$ Mpc ($N_2$, second from the bottom); and redshift (bottom). For comparison, we display $f_{\rm correct}$ for the simple majority in red and for unanimous agreement in blue, accompanied by a dashed black line that marks the average ensemble performance as a guideline.

In the case of both the stellar masses and the gas fractions, the performance of the models is robust throughout, with a slight dip at intermediate values for both, but not significant considering the error-bars. However, in terms of environmental factors, clear trends are observed. The models outperform the average in situations where $r_2 < 70$ kpc. This suggests that galaxies undergoing a merging event, with another close neighbour present, potentially leading to a subsequent merger, are likely experiencing more dramatic encounters, making their features easier to detect. Additionally, $N_2$ is as important as $r_2$, and the performance of the models sinks rapidly when approaching denser environments with $N_2 > 50$. This is evident, as denser environments may have a higher incidence of projections that mimic pair interactions, but are largely rare. Finally, the models are stable with respect to the redshift, dipping below average only at the high end of our redshift domain at $z > 0.2$.

In summary, we do not see significant trends with stellar masses, gas fractions, and redshift, particularly in the regions of parameter space where most galaxies lie. However, it is evident that the environment plays a significant role in the observability of these mergers and their impact on the models' performance.

\begin{figure}
    \centering
    \includegraphics[width=0.45\textwidth]{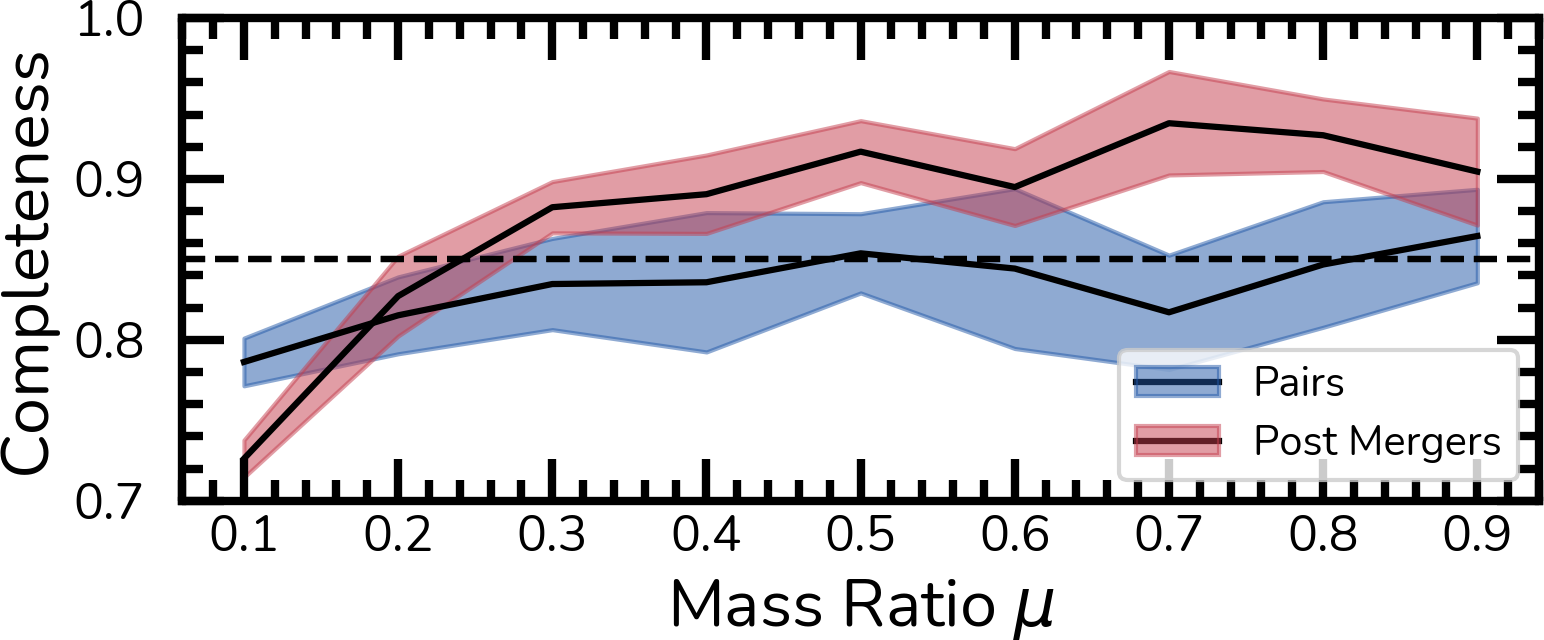}
    \caption{\textbf{Completeness vs. Mass Ratio.} We show the impact of mass ratio on the completeness of galaxy mergers, categorized by stage, within the overall galaxy population in the sample. Pairs are represented in blue and post-mergers in red. For this analysis, we rely on intrinsic labels from the simulations rather than predictions from \texttt{STEP2}.}
    \label{fig:massratio}
\end{figure}

Before moving on to the merger stage classification with \texttt{STEP2} of \textsc{Mummi}, which focuses solely on the merger classes distinct from the overall galaxy population, we first explore whether there are classification biases concerning the mass ratio amongst all galaxies. Given that only galaxy mergers (and not controls) possess a valid mass ratio in the simulations, Fig.~\ref{fig:massratio} examines the influence of mass ratios on the completeness of mergers classified by \texttt{STEP1}. The merger stages are shown, with pairs in blue, and the post-mergers in red. These are based on the intrinsic labels from our mergers selections and not the \texttt{STEP2} predictions.

The results in Fig.~\ref{fig:massratio} agree with the expectation that galaxy mergers with lower mass ratios are harder to detect (minor mergers), with completeness for $\mu < 0.2$ lower than $80\%$ \citep{Wilkinson2024}. However, for higher mass ratios, post-mergers exhibit a higher completeness compared to pairs. The completeness of pairs remains stable, aligning with the mean performance of the ensemble. In cases of major mergers ($\mu > 0.25$), the completeness of post-mergers is approximately $90\%$, regardless of their timescale.

Given evidence that mini mergers significantly enhance galaxies' asymmetric features due to their prolonged timescales and frequency \citep{Bottrell2024}, we investigate the impact of mergers with mass ratios ($0.01 < \mu < 0.1$), below our selection threshold, on the contamination of our classifications. In our test sample, encompassing both mergers and controls, $30\%$ are mini mergers occurring within $1.7$ Gyr. Among the false positives -- comprising solely control galaxies -- the rate of mini mergers increases to $46\%$. This implies that relaxing the mass ratio criteria could reclassify some false positives as mergers, suggesting \textsc{Mummi} detects merging features even in the control sample.

\subsection{\texttt{STEP2}: Merger Stage Classification} \label{subsec:mergerstageclassification}

We now explore the performance of \texttt{STEP2} in our \textsc{Mummi} framework to separate the galaxy mergers into pairs or post-mergers. Given that we expect a similar distribution of pairs and post-mergers in the real universe for a defined timescale, employing methods to address class imbalance is not deemed necessary at this step. We adopt the simple approach of using a single pair of models, a CNN and a SwinTransformer, to classify the stage of the mergers, instead of a large ensemble as in \texttt{STEP1}. 

\begin{figure}
    \centering
    \includegraphics[width=0.45\textwidth]{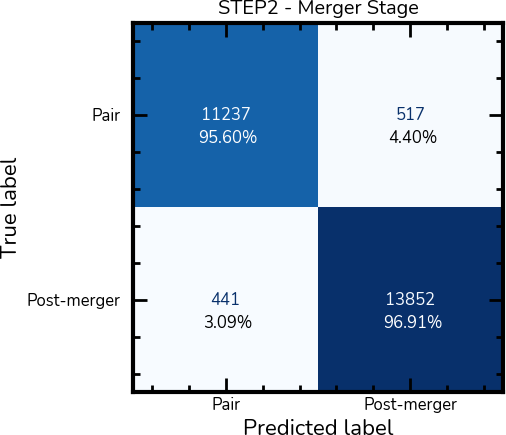}
    \caption{\textbf{Confusion matrix for merger stage classification}. We display the confusion matrix for galaxies in the \texttt{STEP2} test set, assuming that the non-mergers have already been filtered out from the sample with \texttt{STEP1}. There is minimal confusion between the merger stages, with misclassifications accounting to approximately $4\%$. This indicates that with a pure classifier at \texttt{STEP1}, effectively separating mergers into different stages is feasible using a small ensemble or a single model.}
    \label{fig:CM2}
\end{figure}

Here we assume that our classifications in \texttt{STEP1} are accurate and focus solely on evaluating the intrinsic performance of \texttt{STEP2}. Using the original pool of images, with classifications averaged based on redshift, we analyze $26,047$ galaxy mergers. Of these, $11,754$ are pairs, and $14,293$ are post-mergers. Fig.~\ref{fig:CM2} presents the confusion matrix for the merger stage classification. The models achieve $95.6\%$ purity for pairs and $96.9\%$ for post-mergers, indicating minimal confusion between these stages.

This demonstrates that, with a robust method for identifying  mergers in \texttt{STEP1}, we can expect minimal confusion between the pre- and post-merger stages. This approach effectively reduces the false positive rate of pairs mistakenly classified as post-mergers. In this sense, the use of a hierarchical method that includes pairs during training -- as opposed to focusing on a single stage from the outset -- enhances our understanding of how pairs influence post-merger classification. For instance, in \citet{Bickley2021}, a significant number of false positives in post-merger classifications were actually pairs of galaxies. However, as no model was trained to identify these pairs, they could not be filtered out in real-galaxy applications where ground truth is unavailable. Even though pairs can be confirmed spectroscopically, this requires that spectroscopy data is available for the filtering.

The convolved effect of \texttt{STEP1} and \texttt{STEP2} is assessed jointly in \S~\ref{sec:mockresults}. In the observations, we apply bot methods idependently, and only use the \texttt{STEP2} classifications if \texttt{STEP1} voting thresholds are suscessfully met.

\subsection{Mock Survey}\label{sec:mockresults}

To showcase \textsc{Mummi} within a realistic survey context, accounting for the class imbalance issues due to the universe's intrinsic merger rate, we classified all $303,104$ galaxies in the mock survey created from IllustrisTNG-100 using both \texttt{STEP1} and \texttt{STEP2} of our framework. Although this dataset includes galaxies from our training and test sets, their mock images were generated independently with unique viewing angles and at random redshifts for each galaxy, rather than from all realizations (details in \S~\ref{subsec:mocksurvey}).

We label the galaxies in the mock survey more appropriately to align with what is typically encountered in real observations. First, we assign all galaxies with at least one companion with $M_* > 10^9 M_\odot$ within $50$ kpc to the pairs class. From the remaining mergers, any galaxy with a merger event of mass ratio $\mu \geq 0.1$ in its past $1.7$ Gyr is considered a post-merger. The remaining galaxies are all assigned a non-merger label. In summary, the mock survey comprises $48,396$ pairs, $28,835$ post-mergers, and $225,873$ non-mergers, reflecting a 1:5 and 1:10 imbalance. It is important to note, however, that the mock survey includes massive galaxies with $M_* > 10^{11} M\odot$ that are outside of our training sample domain.

\begin{figure}
    \centering
    \includegraphics[width=0.45\textwidth]{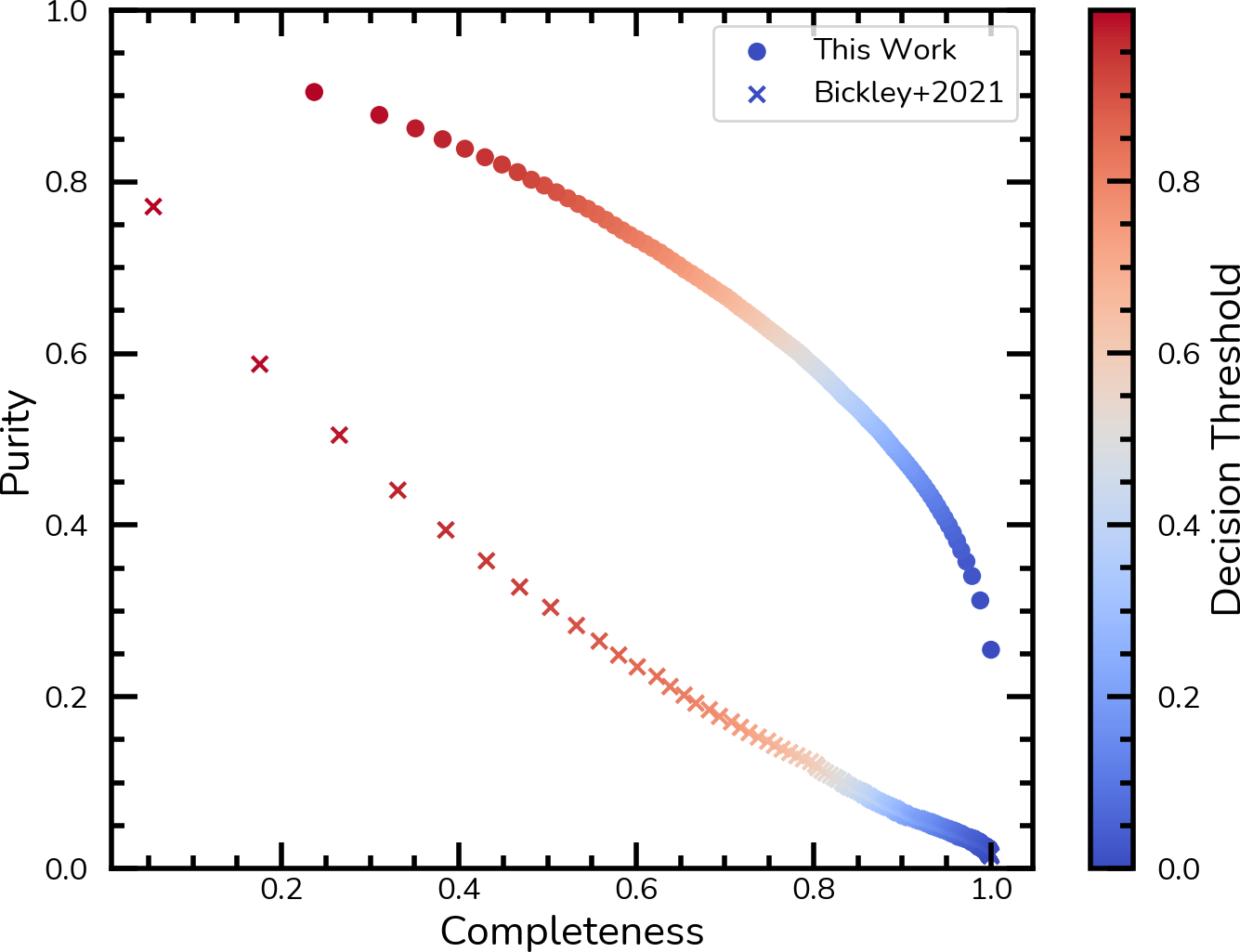}
    \caption{\textbf{Purity-Completeness in the Mock Survey.} We show the purity-completeness trade-off of \textsc{Mummi} in comparison to what was achieved by \citet{Bickley2021}. The decision threshold correspond in our case to the number of votes divided  by the total votes. Our results are shown as circles while crosses denote the results from \citet{Bickley2021}. Although the comparison between the two studies is not direct since the models were trained for slight distinct purposes, our new method has a better purity-completeness tradeoff, such that ultimately it will be more complete and include fewer false positives overall. }
    \label{fig:mocksurveycompleteness}
\end{figure}

A direct comparison with the mock survey results of \citet{Bickley2021}, who used a single network, is not straightforward since we adopt very distinct training strategies, with different timescales and merger stages included. However, we can explore our completeness-purity trade-off in comparison to theirs as it only looks at the overall success of the models in general. To do this, in Fig.~\ref{fig:mocksurveycompleteness} we display the completeness-purity trade-off colour-coded with a divergent colour-map colour coded by the decision threshold. Our results are shown as circles while \citet{Bickley2021} are shown as crosses. In our case, our decision threshold is based on the number of votes dived by the total number of votes possible (20), while for \citet{Bickley2021} it is the decision threshold directly applied to the model probability. For all decision thresholds, our purity and completeness are always higher. For instance, at $30\%$ completeness, we achieve around $90\%$ purity, while in \citet{Bickley2021} is around $50\%$. Fig.~\ref{fig:mocksurveycompleteness} therefore demonstrates the improvement in classification performance in a survey setting using our new \textsc{Mummi} approach compared with the traditional single network method.

We stress that this mock survey approach is the best way to evaluate the performance of \textsc{Mummi} based on an intrinsic merger fraction. As we use all galaxies available for the mass threshold of this study, the merger fractions here are similar to the real Universe.
%A naive conclusion would be that our new method is intrinsically more pure and complete in all cases. However, as we include longer timescales, many of the galaxies that are just outside their CNN classifications in timescale are considered wrong classifications. In our case, on the other hand, these galaxies will show up as correctly classified. Thus, by including a more relaxed merger definition in our training sample, we are able to identify a significantly larger merging galaxy sample. 

\subsection{Merger Candidates in the UNIONS DR5-SDSS DR7 Overlap}\label{sec:postmergersCFIS}

We apply both steps of \textsc{Mummi} to the UNIONS DR5-SDSS DR7 overlap (\S~\ref{subsec:CFIS}) to search for merger candidates in an automated fashion. We generate UNIONS r-band cutouts with $12$ Petrosian radius field of view, and apply the same pixel value normalizations as our mock observations pipeline (\S~\ref{subsec:realsim}). This is a slight difference from the IllustrisTNG cutouts used to train the networks, but necessary given we want to apply \textsc{Mummi} to lower and higher stellar mass galaxies than what is possible with out training set, anticipating generalization to lower and higher masses, but restricting our analysis to redshift ranges that align with our mock pipeline.

%Using SDSS spectroscopic redshifts, 

From $235,354$ sources in our UNIONS DR5-SDSS DR7 overlap, $42,763$ ($\sim18\%$) reached a simple majority vote (11 votes) for merger candidacy, while $13,448$ ($\sim5\%$) received an unanimous merger classification (20 votes). 

\begin{figure*}
    \centering
    \includegraphics[width=1\textwidth]{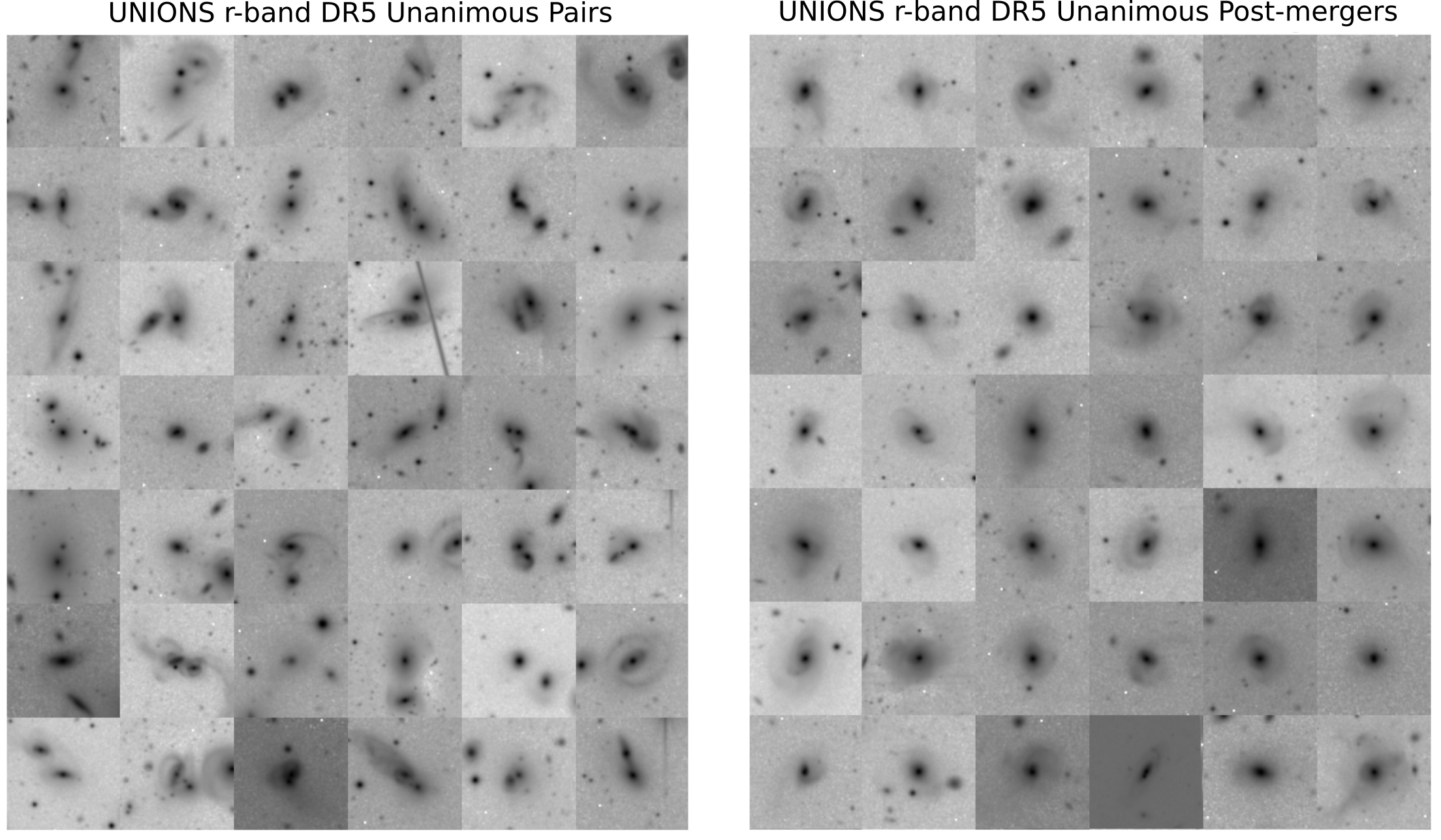}
    \caption{\textbf{Examples of merger candidates in UNIONS DR5 with \textsc{Mummi}}. We show 42 randomly selected pairs (left) and 42 post-mergers (right) from the unanimous sample of merger candidates in the UNIONS DR5-SDSS DR7 overlap sample. All displayed galaxies unanimously received 20 votes in \texttt{STEP1}, with \texttt{STEP2} further categorizing them into the two panels shown. The pairs predominantly feature interacting morphologies with multiple sources, often accompanied by tidal features. Conversely, post-mergers are generally isolated galaxies exhibiting clear distortions in their light profiles, including tidal features, shells, and warped spiral arms.}
    \label{fig:cfisexample}
\end{figure*}

Opting for an automated approach without a visual veto, we focus exclusively on unanimous cases to determine their merger stage using \texttt{STEP 2} of \textsc{Mummi}, considering these as the strongest candidates. Of the $13,448$ merger candidates, $9,514$ were classified as pairs/pre-mergers, and $3,934$ as post-mergers. In Fig~\ref{fig:cfisexample} we show examples randomly drawn from the samples of unanimous pairs (left) and post-mergers (right) in r-band of UNIONS. The visual morphology of pairs is dominated by the presence of multiple galaxies. However, many exhibit very clear tidal features. Conversely, post-mergers are typically isolated, displaying evident disturbances in their morphologies. The images in Fig.~\ref{fig:cfisexample} thus provide confidence in \textsc{Mummi}'s performance in classifying mergers.

The catalog with \textsc{Mummi} classifications in the UNIONS DR5-SDSS DR7 overlap is released alongside this work\footnote{Catalogs are going to be publicly available at \url{https://github.com/astroferreira/MUMMI_UNIONS} after acceptance.}. While we recommend selecting candidate mergers based on unanimous model agreement for purely automated classifications, the catalog also contains details on \texttt{STEP1} voting and \texttt{STEP2} probabilities. This approach enables readers to make their own selections and fully leverage the framework's capabilities for their specific research needs.

A complete analysis of galaxy mergers in UNIONS is left to future work where we will apply \textsc{Mummi} to the whole area covered by UNIONS, not only on the overlap with SDSS, expanding to a photometric redshift based sample.

\section{Discussion}\label{sec5-discussion}

\subsection{A bridge between cosmological simulations and observations}

Simulation-driven methods rely on approximations of the real universe, and depend on the accuracy of mathematical models used to describe reality for their success. The underlying physics used to model dynamical friction, interactions, and the evolution of galaxy mergers are fairly well understood. The morphologies produced by simulations, such as IllustrisTNG, are realistic enough to replicate the set of observables seen in the real universe \citep{Zanisi2021}. Additionally, IllustrisTNG in particular has been tested and extensively used to study galaxy interactions and mergers \citep{Patton2020, Hani2020, Salvatore2021, Byrne-Mamahit2023, Brown2023, Byrne-MamahitB, Patton2024}.

However, it is crucial to introduce realistic instrumental and observational features into these simulated images \citep{Bottrell2019}. A formalism that does not account for these features can lead to catastrophic results when applied to real observations. \citet{Alexandra2020, Alexandra2021} demonstrates that when image characteristics are different between two domains, domain adaptation techniques might be necessary to improve performance. Thus, it is not enough to train models on idealized simulations and apply them directly to real galaxies without assessing if the features used by these models are in fact similar.

To address this, several works investigate the representations generated by deep learning models both in the simulation domain and in the observational domain. For example, \citet{Ferreira2022} demonstrate that CNNs trained with IllustrisTNG generate similar representations for both simulated galaxies and galaxies observed in the CANDELS fields. More recently, self-supervised approaches with contrastive learning were shown to generate powerful feature spaces even in the absence of labels during training, by focusing on the similarities and the differences between the domains \citep{huertas2023, ERGOML}. Although these methods effectively create a bridge between the domains, they incorporate observational data into the training framework, thereby conditioning the models to it a priori.

Here we will test if it is possible to create representations that can connect both domains with supervised learning alone, without including any real data during the training phase. To do this, we create a feature extractor from our already trained SwinTransformer models in \texttt{STEP1}, extracting their attention maps before the classification head. This amounts to $12,288$ total features ($8\times8\times192$) per galaxy\footnote{We only use the SwinTs because the EfficientNet representation space is even higher dimension, and the resulting representations were less intuitive.}. Subsequently, we apply these models to the $303,104$ galaxies in the mock survey and $235,354$ real galaxies in the UNIONS DR5 sample that have a matching a spectroscopic redshift for a source in SDSS. As twenty models were used per galaxy, there are twenty representations each. We average these features out into a single 192 length vector per galaxy. From these features, we then create two Unified Manifold Approximation and Projections \citep[UMAPs, ][]{UMAP}, one for the IllustrisTNG100-1 data and one for CFIS DR5 galaxies.

\begin{figure*}
    \centering
    \includegraphics[width=1\textwidth]{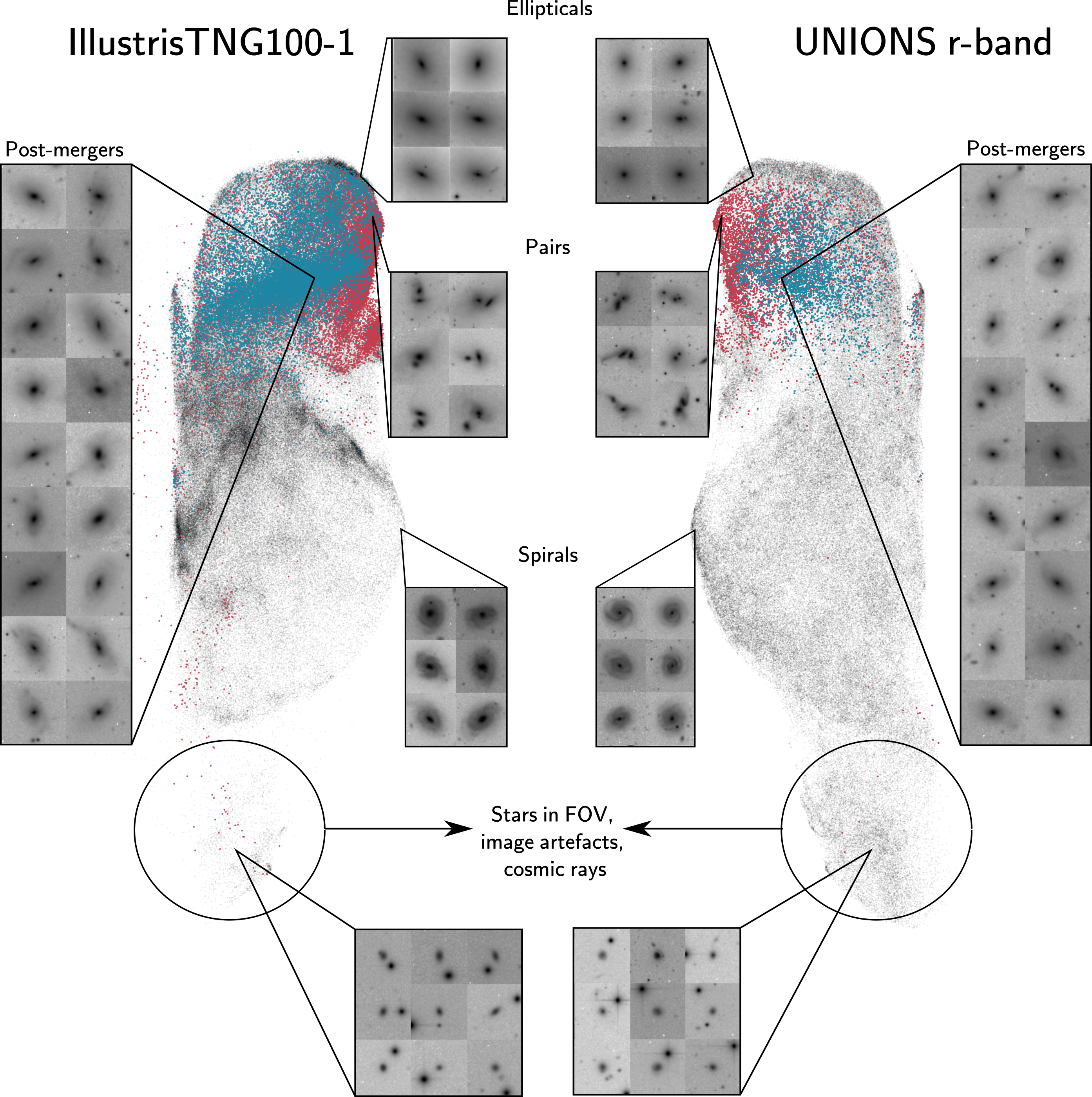}
    \caption{\textbf{Feature representations for IllustrisTNG100-1 (left) and UNIONS (right).} We show the representations generated by UMAPs on the attention features extracted from the SwinTransformers for both the simulations and observations. The UMAPs organize the morphologies in a two-dimensional space such that we can search regions of this space symmetrically between the both domains to extract galaxies that share similarities. We mirrored the UMAP of IllustrisTNG in the x-axis so that we can group the morphologies symmetrically. We colour code by their merger stage, with \textbf{pairs in red and post-mergers in blue}, respectively. Black dots are galaxies not classified as mergers by our ensemble of models. We demonstrate that even though these representations are not perfectly equal, the morphologies of interest are distributed in the same fashion, such that the cluster of post-mergers in the simulations lies in the same region on the observations. Additionally, it is shown that these representations not only organize the merging morphologies but other, more general classes, such as spirals and ellipticals. }
    \label{fig:umaps}
\end{figure*}

In Fig.~\ref{fig:umaps} we show both unbinned UMAPs, with the IllustrisTNG mock survey on the left and the UNIONS DR5 in the right. Note that the UMAPs in Fig.~\ref{fig:umaps} are mirrored for visualization purposes. Each non-merger galaxy is displayed as a black dot, while the pairs are shown in red and the post-mergers in blue. To explore this representation space, we created a tool that performs a simultaneous nearest neighbours lookup in both UMAPs. If a data point is selected in the IllustrisTNG100-1, we can find its closest neighbours in the UNIONS space and vice versa. This serves to determine if the morphological arrangement is very similar between the domains, which we indeed find to be the case, as shown by the example galaxies in Fig \ref{fig:umaps}. We highlight the exact same region in both diagrams for some key morphologies, such as elliptical galaxies (top), pairs (middle), spirals (second from the bottom), and post-mergers in the external larger pane. Images with artefacts, gaps between observations, large diffraction spikes from luminous stars, and cosmic rays are concentrated at the bottom region of the UMAPs within the circles. Notably, these features are found in both the real UNIONS data and in the simulation data with realism added. For example, this parameter space can be used to quickly  filter out stamps with artefacts in a clean way as these regions have no overlap.

Both UMAP representations show that models trained using supervised learning are capable of going beyond classification given enough data for proper generalization. Here, the specific arrangement of the morphologies is important, as the merging galaxies lie between the disks in the middle and the spheroids on top, creating a sort of physically motivated Hubble tuning fork \citep{Hubble1926}. Note that the only labels that these models were trained to predict were if the galaxies had merger events or not. 

Another notable feature of these representations is the arrangement from left to right, organizing galaxies from structured to structure-less. For example, spirals are found in the lower left region of the UMAP, while featureless disks are spread out to the right. By searching this parameter space, pockets of similar morphologies can be defined, such as galaxies containing bars and rings. Moreover, specific galaxies in the observations that have a particular morphology can be traced  to the same region in IllustrisTNG-100, creating a direct way to link real morphologies to simulated morphologies.

\begin{figure*}
    \centering
    \includegraphics[width=1\textwidth]{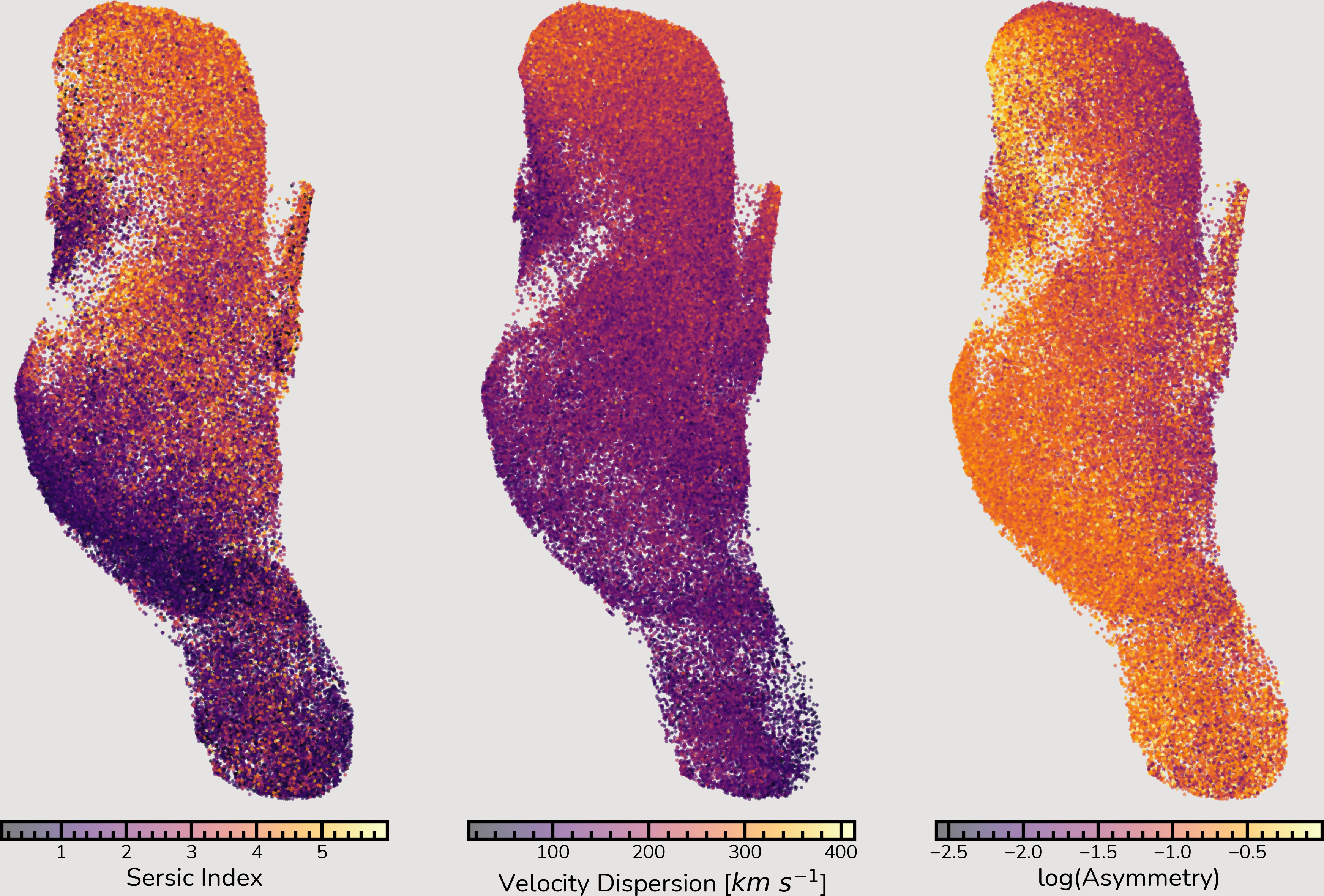}
    \caption{\textbf{Properties correlated with UMAP representations in UNIONS}. We show the correlation of Sersic indexes (left), velocity dispersions (middle), and asymmetries (right) to the UMAP's spatial representations. Galaxies exhibit steeper light profiles and higher velocity dispersions as they approach the merger region (see Fig.~\ref{fig:umaps}). Similarly, asymmetries vary across the representation space, with structured (or disturbed) galaxies predominating on the left, and galaxies with low asymmetry clustering toward the right edge. Furthermore, the regions with the highest asymmetry align with the merger area. These correlations showcase the powerful nature of the UMAP representations. Note that only galaxies with valid measurements are shown in each panel, which explains any slight variations observed between the points distribution.}
    \label{fig:nugget}
\end{figure*}

To further illustrate that the UMAP representation is organized in a powerful way to probe merger driven galaxy evolution between simulations and observations, in Fig.~\ref{fig:nugget} we explore correlations of the UNIONS 2D space with the Sersic index \citep[parametric morphology, ][]{sersic1963}, asymmetries \citep[non-parametric morphology,][]{Abraham1996, Conselice2003a} and velocity dispersion \citep[kinematics,][]{Cappellari2011, Bluck2019}. We measure the morphology indices running the \textsc{Morfometryka} code \citep{Ferrari2015} on the UNIONS r-band, and use the velocity dispersion information available from SDSS DR7 spectra \citep{Abazajian2009}. 

The correlation between the three properties displayed in Fig~\ref{fig:nugget} and the UMAP representations reinforce the role of merging to galaxy evolution. As we go from the bottom to the top, and through the region where the galaxy mergers are preferentially situated (see Fig.~\ref{fig:umaps}), galaxies display steeper light profiles with higher Sersic indices (left panel). The velocity dispersion in the middle panel also highlights that galaxies under the merger region have low velocity dispersions (and thus are disc-dominated), whereas galaxies above it have the highest velocity dispersions, showing that these systems are not rotationally supported, likely galaxies that underwent subsequent merging events. Furthermore, the asymmetries in the right panel show that galaxies are also organized from left to right based on their structure. Most low asymmetry galaxies are located at the right edge of the representations, whereas the sources in the left have considerably higher asymmetries. Moreover, the galaxies with the highest asymmetries are all clustered in the merger dominated region, especially around the pair area.

A complete description of these representations and their potential applications is beyond the scope of this paper. A future paper will delve into our results within UNIONS DR5, discussing how we can take \textsc{Mummi} one step further to measure merger properties that are otherwise unobtainable in observations, such as merger timescales.

\subsection{Model Biases On UNIONS Observations}

While training models on synthetic data, one issue that arises is the impact of limitations in the training data transferring over to the inference step with real observations, resulting in biases. In our case, for example, our training set is limited in stellar mass ($10^{10} < M_*/M_\odot < 10^{11}$), while we apply \textsc{Mummi} to a wider domain. Understanding any offsets between the overall distribution of galaxies we want to apply \textsc{Mummi} to and the resulting distribution from the classifications reveals some of these potential biases.

\begin{figure*}
    \centering
    \includegraphics[width=1\textwidth]{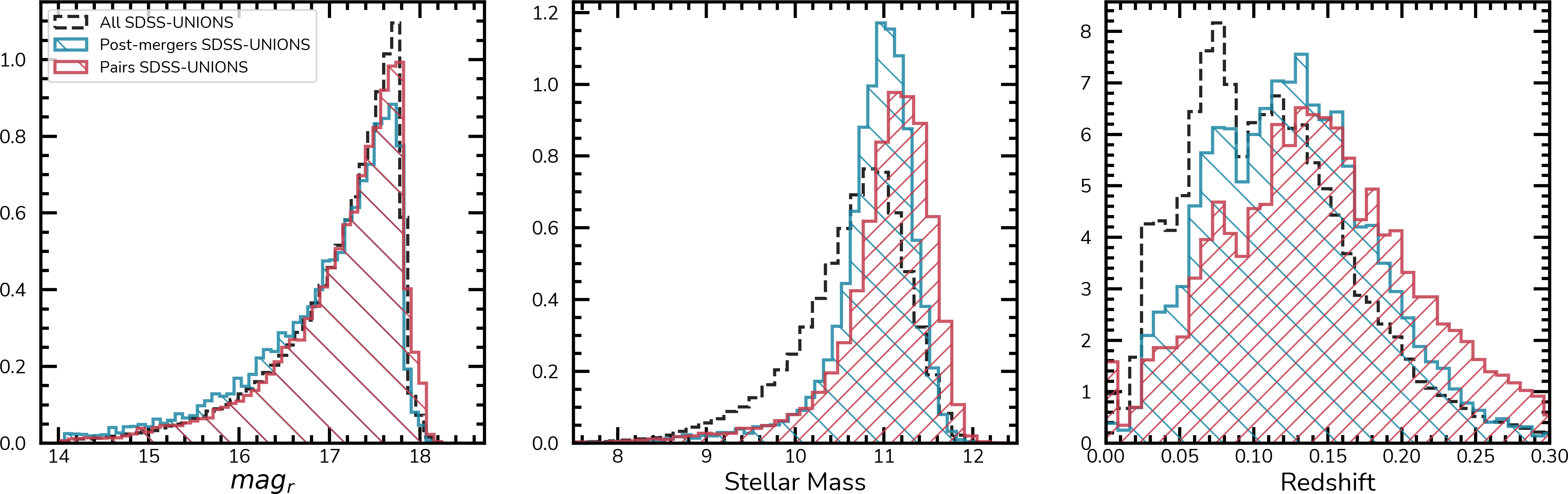}
    \caption{\textbf{Classification biases in the SDSS-UNIONS overlap.} We display the distributions of r-band petrosian magnitude (mag$_r$), stellar mass and redshift for the whole SDSS-UNIONS distribution (black) and the \textsc{Mummi} classifications. Post-mergers are shown in blue while pairs are shown in red. The \textsc{Mummi} classifications are in general biased towards higher stellar masses and higher redshifts than the overall sample. However, there is no brightness bias, as can be seen by the distributions almost perfect overlapping in mag$_r$. These trends can be directly compared to the work in \citet{Bickley2022}, where the offsets in all three properties are more severe.}
    \label{fig:biases}
\end{figure*}

In Fig.~\ref{fig:biases} we show distributions of r-band Petrosian magnitudes (left), stellar mass (centre) and redshift (right), for the whole SDSS-UNIONS sample (black dashed line) and for post-mergers and pairs from \textsc{Mummi} classifications in blue and red, respectively. As can be seen, there is no apparent brightness bias, as the distributions track each other well. On the other hand, we find important biases in stellar mass and redshift. 
Both pairs and post-mergers display higher stellar masses  than the available distribution. However, this offset is expected, as the model never sees central galaxies with masses $< 10^{10} M_\odot$. For example, a similar trend is found by \citet[][see their Fig.3]{Bickley2022}. Nevertheless, we find a higher fraction of lower mass galaxies to be mergers compared to \citet{Bickley2022}, indicating that our models achieve better out-of-distribution success. Mitigating stellar mass biases will require the use of higher resolution simulations, such as TNG50-1 or galaxy-galaxy zoom-ins, to include lower mass galaxies in the training set.

In the case of redshifts, we also find a small offset in the classifications, with a higher redshift on average for the mergers when compared to the whole distribution. This can be attributed to both the training set and also the fact that images need to be re-scaled to the model input size. In our training sample, the lower redshift bin used is $z=0.015$, and only 3 redshift realizations probe $z<0.05$. This might favour the models towards $z>0.05$, as most of the training data is at that range. Additionally, low redshift galaxies span a larger intrinsic angular field of view, including more pixels (and thus information), but they still need to be re-binned to the model input size of $256 \times 256$. Unfortunately, this limits the capabilities of the models to leverage all information available in high resolution low redshift images. Training models with extremely high resolution input images and up-scaling from high redshift images is impractical at this stage due to the computational resources required. Nevertheless, the models presented show a smaller bias in redshift in comparison to \citet{Bickley2022}. Knowing the extent of these biases aids us in understanding potential sources of incompleteness in our classifications with the real observations.

\section{Conclusion}\label{sec6-conclusion}

In this paper, we present \textsc{Mummi} (Multi Model Merger Identifier), a simulation-based multi-step hybrid CNN-Vision Transformer framework for classifying galaxy mergers in the r-band imaging of UNIONS. \textsc{Mummi} is able to identify and characterize the stage of merging galaxies, even at long timescales of up to $1.75$ Gyr before or after a merging event. To train \textsc{Mummi}, we generated a 6.4 million image synthetic dataset from the IllustrisTNG100-1 simulation that has UNIONS realism-added properties using \textsc{RealSimCFIS}. Additionally, to address the rarity of galaxy mergers in the real universe, \texttt{STEP1} of our framework employs a large ensemble of models combined through a jury-based voting system. Then, \texttt{STEP2} of the framework focuses on the classification of the merger stage, separating pairs and post-mergers. The main takeway points of our work are:

\begin{itemize}

    \item \textbf{\textsc{Mummi} selects pure samples of mergers irrespective of their timescales}. Using the ensemble of models in \texttt{STEP1} in a jury-based decision making, where only galaxy mergers where models unanimously agree on are selected, produces merger samples with purities above $95\%$. This mitigates the false positive rates that naturally arise from the imbalance of galaxy mergers among the overall population of galaxies in the real universe (Figs \ref{fig:timescaleperformance}-\ref{fig:purityvsensemblesize}).
       
    \item \textbf{\texttt{STEP1} models are robust with respect to stellar masses, gas fractions, mass ratios and redshifts}. \textsc{Mummi} identifies pairs and post-mergers achieving $84\%$ accuracy, $85\%$ purity, and $82.5\%$ completeness when used with a simple majority for galaxy mergers in any merger stage within a $\pm1.75$ Gyr of the merging event. However, they are impacted by environment, especially in crowded systems (Fig.~\ref{fig:environmentstats}-\ref{fig:massratio}).
    
    \item \textbf{\textsc{Mummi} successfully separates pairs and post-mergers without human intervention}. \texttt{STEP2} in our framework is able to characterize the stage of a merger (pre/pair or post-merger) at a $96\%$ success rate, further reducing the false positive rates produced by pair of galaxies in automated post-merger selections (Fig.~\ref{fig:CM2}). 

    \item \textbf{\textsc{Mummi} excels when applied to imbalanced datasets compared to other works}. We showcase the full power framework in a mock survey, demonstrating how our large ensemble of models can be used together to reliably produce purer samples in the survey context. The false positive rates of post-mergers drop by $75\%$ while the completeness lost is less than $50\%$. Additionally, \textsc{Mummi} have better completeness-purity trade-offs when compared to previous works (Fig. \ref{fig:mocksurveycompleteness}).

    \item \textbf{We report a high-confidence sample of $13,448$  galaxy mergers ($9,514$ pairs, $3,934$ post-mergers) in the UNIONS DR5-SDSS DR7 spectroscopic overlap}. We publicly release a catalog with classifications to all $235,354$ sources in the UNIONS DR5-SDSS DR7 overlap. We find $42,763$ merger candidates. From these $13,448$ received unanimous classifications (20 votes). Alongside this catalog, we report all voting information and class probabilities to enable the user to generate their own selections (Fig.~\ref{fig:cfisexample}). 

    \item \textbf{We show that the representations created by our models can act as a powerful bridge between  simulations and observations using a simple supervised training approach.} By using features extracted from simulated galaxies and real galaxies in UNIONS DR5. The representations organize the morphologies in a such a way to create a physically motivated Hubble tuning fork, where galaxy mergers are placed as a transitioning region between disk galaxies and spheroids (Fig.~\ref{fig:umaps}-\ref{fig:nugget}).
\end{itemize}

In this paper, we showcased \textsc{Mummi} mainly in the IllustrisTNG context, discussing the details of our implementation. In an upcoming paper, we will discuss the results of \textsc{Mummi} applied to UNIONS DR5 in more detail. One of our primary objectives is to predict timescale information in real observed galaxies, enabling us to carry out simulation-like temporal characterization of galaxies in the merger sequence using real-Universe observations.

\section*{Acknowledgements}
% Entry for the table of contents, for this guide only
\addcontentsline{toc}{section}{Acknowledgements}
We are honored and grateful for the opportunity of observing the Universe from Maunakea and Haleakala, which both have cultural, historical and natural significance in Hawaii. This work is based on data obtained as part of the Canada-France Imaging Survey, a CFHT large program of the National Research Council of Canada and the French Centre National de la Recherche Scientifique. Based on observations obtained with MegaPrime/MegaCam, a joint project of CFHT and CEA Saclay, at the Canada-France-Hawaii Telescope (CFHT) which is operated by the National Research Council (NRC) of Canada, the Institut National des Science de l’Univers (INSU) of the Centre National de la Recherche Scientifique (CNRS) of France, and the University of Hawaii. This research used the facilities of the Canadian Astronomy Data Centre operated by the National Research Council of Canada with the support of the Canadian Space Agency. This research is based in part on data collected at Subaru Telescope, which is operated by the National Astronomical Observatory of Japan.
Pan-STARRS is a project of the Institute for Astronomy of the University of Hawaii, and is supported by the NASA SSO Near Earth Observation Program under grants 80NSSC18K0971, NNX14AM74G, NNX12AR65G, NNX13AQ47G, NNX08AR22G, 80NSSC21K1572 and by the State of Hawaii.
We acknowledge the support of the Digital Research Alliance of Canada for providing the compute infrastructure to train the models presented here. We thank Sarah Huber for helpful technical support on the usage of the Cedar cluster, troubleshooting and helping with optimizations necessary to manage the data scale of our models and datasets swiftly. We acknowledge and thank the IllustrisTNG collaboration for
providing public access to data from the TNG simulations. We thank Mike Hudson and Florence Durret for discussions regarding the content of this paper. SLE, DRP and LF gratefully acknowledge NSERC of Canada for Discovery Grants which helped to fund this research. SJB and SW acknowledge graduate fellowships funding from the Natural Sciences and Engineering Research Council of Canada (NSERC); Cette recherche a été financée par le Conseil de recherches en sciences naturelles et en génie du Canada (CRSNG).

%%%%%%%%%%%%%%%%%%%%%%%%%%%%%%%%%%%%%%%%%%%%%%%%%%

\section*{Data Availability}

A subset of the raw data underlying this article are publicly available via the Canadian Astronomical Data Center at http://www.cadc-ccda.hia-iha.nrc-cnrc.gc.ca/en/megapipe/. The remaining raw data and all processed data are available to members of the Canadian and French communities via reasonable requests to the principal investigators of the Canada-France Imaging Survey, Alan McConnachie and Jean-Charles Cuillandre. All data will be publicly available to the international community at the end of the proprietary period, scheduled for 2023.



%\bibliographystyle{mnras}

\bibliographystyle{mnras}
\bibliography{example}

%%%%%%%%%%%%%%%%%%%%%%%%%%%%%%%%%%%%%%%%%%%%%%%%%%

%%%%%%%%%%%%%%%%% APPENDICES %%%%%%%%%%%%%%%%%%%%%

%%%%%%%%%%%%%%%%%%%%%%%%%%%%%%%%%%%%%%%%%%%%%%%%%%
\appendix

\section{Training Setup}\label{subsec3:training}

In this appendix we briefly describe our training setup, including the hardware used, training configuration, and specific details to enable the reproduction of this experiment.

Each individual model is trained on 4xV100-32GB GPUs, utilizing a batch size of 2048, until the model converges. A total of 22 models were trained, 20 for the merger identification ensemble and 2 for the merger stage classification task. Model training averages $\sim16$ hours per model. To prevent disk readout bottlenecks, our dataset is stored in the binary \texttt{tf.data.TFRecords} format, segmented into 8192 image shards. Each shard contains images of 128 by 128 pixels, normalized to a range between 0 and 1.

To regularize our model training, we add a custom on-the-fly data augmentation pipeline that introduces the augmentation transformations directly to the loading pipeline graph using the \texttt{.map()} function from \texttt{tf.data}. Our pipeline comprises seven augmentation steps, each with a $70\%$ probability of being applied. These include random vertical and horizontal flips; random 90-degree rotations; small-angle rotations (-10 to 10 degrees); positional shifts of up to 5 pixels; shear distortions up to $15\%$ on both axes; zooms of up to $30\%$ of the image size; and occlusion augmentations. Each of these is introduced to increase the variability of the morphologies across different epochs and to decrease overfitting. The occlusion augmentation specifically acts by preventing the model from overly relying on a few specific features in the images, thereby forcing it to classify them even when parts are missing. 

\begin{figure}
    \centering
    \includegraphics[width=0.48\textwidth]{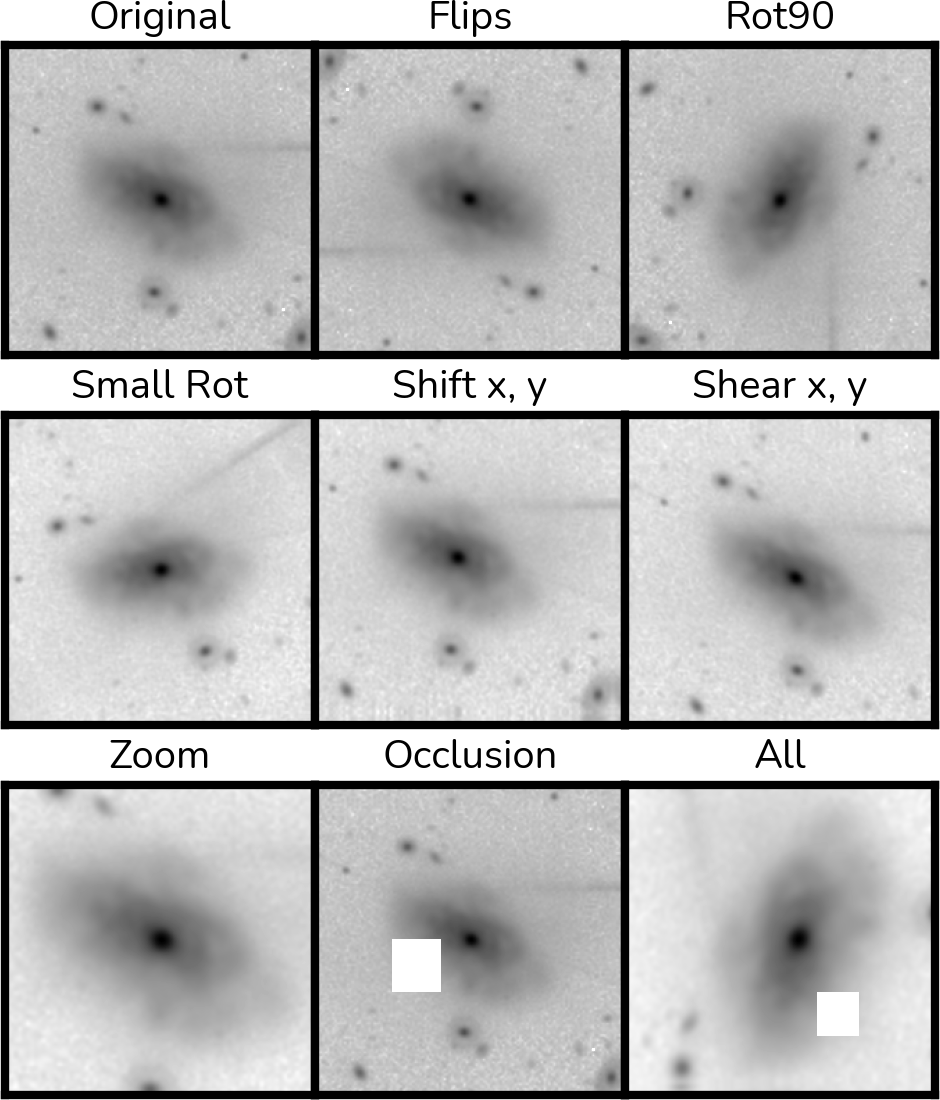}
    \caption{\textbf{Data augmentation pipeline.} We show all possible steps in our on-the-fly data augmentation pipeline, starting from the original image (top left), we apply each augmentation with a $70\%$ probability, including random vertical and horizontal flips, 90 degrees rotation, small angle rotations, centre shifts, image shear distortion in both axis, zoom, and occlusions. The last panel (bottom right) show an example of the final result of all these augmentations are applied together. In practice, however, the augmentations applied will depend on the probabilities, making each image used by the training unique. }
    \label{fig:augmentations}
\end{figure}

In Figure~\ref{fig:augmentations}, we show an example of an original image, each of these transformations, and the combined result of all of them in the last panel. Due to the probabilistic nature and the random scale of each transformation, it is rare for the same image to be transformed in a similar fashion across training epochs. For instance, in each epoch, approximately $9\%$ of the images are not augmented, while another $9\%$ undergo all the augmentations simultaneously. Additionally, in line with the findings of \cite{Walmsley2023}, who demonstrate that deep learning models achieve higher performance with upscaled images, we upscale our images to 256x256 on the fly prior to applying the augmentations.

All models are trained using the \texttt{Lion} optimizer \citep{LionOpt}. We employ a warm-up restart cosine annealing learning rate schedule, oscillating between $10^{-4}$ and $10^{-5}$ over 200 epoch cycles, including a 10 epoch warm-up period. Most models achieve convergence in a single cycle, with a few exceptions. In the merger stage identification step, each model is trained with $499,512$ images and evaluated on $1,248,800$ images. Conversely, in the merger stage classification step, models are trained on $2,924,720$ images and evaluated using $731,280$ images.

% Don't change these lines
\bsp	% typesetting comment
\label{lastpage}
\end{document}